\documentclass[11pt]{article}
\usepackage[T1]{fontenc}
\usepackage[utf8]{inputenc}
\usepackage[margin=1.1in]{geometry}
\usepackage{amsmath,amssymb,amsfonts}
\usepackage{newpxtext}
\usepackage[varg,bigdelims]{newpxmath}
\usepackage{microtype}
\usepackage{graphicx}
\usepackage{textcomp}
\usepackage{xcolor}
\usepackage{comment}
\usepackage{url}
\usepackage{enumitem}
\usepackage{subcaption}
\usepackage{booktabs}
\usepackage{longtable}
\usepackage{algorithm}
\usepackage{algpseudocode}
\usepackage{parskip}
\usepackage[numbers,sort&compress]{natbib}
\makeatletter
\renewcommand\paragraph{\@startsection{paragraph}{4}{\z@}%
  {1.0ex \@plus .4ex \@minus .2ex}{-0.8em}%
  {\normalfont\normalsize\bfseries}}
\makeatother

\title{\textbf{Dense Matrices Are Alike;\\[2pt]
Sparse Matrices Are Sparse in Their Own Way:\\[2pt]A Structure-Adaptive Tile Cholesky Factorization}\\[7pt]}

\author{Esmail Abdul Fattah, Hatem Ltaief, H{\aa}vard Rue, David E. Keyes\\
King Abdullah University of Science and Technology}

\date{}

\begin{document}
\maketitle

\begin{abstract}
Sparse direct Cholesky solvers generally fix one data structure for an entire matrix,
but symmetric positive definite systems in practice range from nearly dense to irregular, with sometimes a mixture of both
within a single matrix. We therefore let the data structure follow the
sparsity structure, both across matrices and across tiles within a single
matrix. Before any numerical work starts, a lightweight selector captures the
sparsity pattern of the Cholesky factor and routes the matrix, on one static
shared-memory schedule, to one of three regimes: dense, sparse, or an
intermediate \emph{semisparse} regime. Dense tiles are stored in full; the
semisparse regime uses a new \emph{active-column tile} that keeps only the
columns the factorization will touch, so structures that are neither dense nor irregularly sparse,
such as banded or arrowhead-shaped matrices common in spatial models,
still reach BLAS-3 efficiency. This spectrum, spanning dense, sparse, and semisparse structure often
within a single problem, is suited to the integrated nested Laplace
approximation (INLA): one sparsity pattern factorized thousands
of times with different values, so the one-time analysis cost is
amortized and every factorization saving compounds. We evaluate the technique on $60$ SPD matrices from Bayesian hierarchical models,
finite-element analysis, and structured grid problems, comparing against
MUMPS, PaStiX, CHOLMOD, symPACK, and Intel oneMKL PARDISO on Intel Xeon
and AMD EPYC nodes; the selector alone achieves the lowest total
factorization time in every regime.
Summed over the suite, it beats the best fixed single-structure mode by
$1.6$ to $2.6\times$, and every alternative by $1.8$ to $12.5\times$ on Intel
and $2.6$ to $10.1\times$ on AMD, with the largest gains on the most
expensive factorizations. It trades more one-time analysis for less time per
factorization, pulling ahead by the third factorization of a given pattern. A first extension of the dense route to one NVIDIA A100 GPU, with the factor kept resident on the device, runs $1.2$ to $6.3\times$ faster than the same route on the faster of the two CPU nodes, the margin widening with the size of the factor. Solver, Python/R/Julia interfaces, benchmark suite, and results are open at
\url{https://github.com/esmail-abdulfattah/sTiles}.
\end{abstract}

\section{Introduction}\label{sec:intro}

Sparse symmetric positive definite (SPD) linear systems are central to
finite-element analysis, interior-point optimization, Kalman filtering, and Bayesian inference for spatial statistics~\cite{davis2016survey,duff2017direct,rue2009approximate}. In these settings, the solver is rarely called just once. Newton steps,
interior-point iterations, time stepping, and hyperparameter sweeps all
re-factor the \emph{same sparsity pattern} while changing only its numerical values: tens of times in a short optimization loop, and hundreds to thousands of times in a sampling or inference run. The workload that motivates this paper is integrated nested Laplace approximation (INLA)~\cite{rue2009approximate}. INLA fits a latent Gaussian model by re-factoring a single sparse precision matrix (the inverse of the covariance) at every hyperparameter
evaluation, so one fit issues thousands of factorizations of the same pattern, many of them independent~\cite{gaedkemerzhauser2023parallelized,abdul2025inla+}.

This kind of reuse changes what a solver should optimize for: the analysis step (ordering, symbolic factorization, and task-graph
construction) is computed once and reused on every call, so a solver for this regime should invest heavily in that one-time analysis and schedule, then replay them on every factorization. The systems involved fit in a single node's memory, so we target shared memory deliberately, without the distributed-memory parallelism across nodes that problems of this size do not require.

Numerical software offers two answers to a linear system, sparse or dense,
and treats the choice as a fixed property of the matrix. But \emph{sparse}
covers an enormous range. Across the suite of Section~\ref{sec:patterns} it runs from densities of $0.0005\%$ to $22\%$ and fill ratios of $1\times$ to $78\times$, and the best treatment at one end of that range differs from the other end as much as either differs from the dense case. The structured middle, dense enough that dense kernels pay off yet far too sparse to store as dense, has no recognized class and no standard representation. The sparse-or-dense label can even be ambiguous within a single matrix. A
handful of variables coupled to everything else, a shared covariate or a
global intercept, completely fill a few rows and columns, while the rest of
the matrix stays nearly empty, so no global choice suits the whole of it.
The right representation therefore varies at two scales: across matrices
and across the tiles of a single matrix.

We present a structure-adaptive tile Cholesky factorization whose unit of
adaptation is the individual tile. On a single fixed grid of $n_b \times n_b$ tiles, each tile takes the form its own fill calls for: dense where the tile is
full, and an \emph{active-column tile} (storing only its active columns)
where it is sparse. At the irregular, ultra-sparse end, where no fixed grid pays off, a non-uniform supernodal tiling runs on the same schedule. A per-matrix selector reads three inexpensive features off the symbolic factor and places each matrix on this spectrum, a decision made once per pattern rather than repaid at every factorization. Our contribution is the system as a whole: a single solver that spans the sparse-to-dense range on one static schedule, built once and replayed across the repeated factorizations. Sparse-direct solvers let the matrix dictate the partition and absorb the resulting irregularity in a dynamic runtime; this design instead fixes the grid in advance. The contributions are as follows.
\begin{enumerate}[leftmargin=1.5em,topsep=2pt,itemsep=2pt]
\item \textbf{A structure-adaptive tile factorization.}
    Using three features read inexpensively off the symbolic factor, a per-matrix selector routes each matrix to one of three tilings (a plain dense tiling, the active-column tile, or a non-uniform supernodal tiling), all sharing one fill-reducing ordering and one static, shared-memory schedule, built once per pattern. The supernodal route reuses an established sparse Cholesky on this static schedule, as a first-class mode rather than a separate solver, and the fixed-grid routes generalize sTiles, which handled only arrowhead matrices, to arbitrary structured sparse SPD systems.
\item \textbf{The active-column tile.}
  A per-tile representation storing only a tile's \emph{active} columns, via
  an active-column mask, an exact-$L$ CSC fill descriptor, and a
  banded-diagonal bandwidth; a dense tile is its fully active limit. The four
  Cholesky kernels run on it as BLAS-3~\cite{dongarra1990blas3} calls with
  symbolically precomputed scatter maps. It suits matrices banded along the
  diagonal, sometimes with a dense strip along one edge, where a tile's
  active columns pack contiguously (Section~\ref{sec:grid-alignment}). Such
  matrices are a fraction of a percent of the suite total, but under repeated
  factorization they are where per-fit time accumulates, so a suite total
  understates the tile (Section~\ref{sec:experiments}).
\item \textbf{A comprehensive experimental study} on $60$ sparse SPD matrices from the
  SuiteSparse collection and from INLA-generated graphs, comparing against
  MUMPS~\cite{amestoy2001fully}, PaStiX~\cite{henon2002pastix}, PARDISO~\cite{onemkl_pardiso}, CHOLMOD~\cite{chen2008algorithm}, and symPACK~\cite{jacquelin2016sympack} on factorization time and
  parallel scalability.
\end{enumerate}

\paragraph{Scope.} We evaluate across the
structural spectrum up to roughly $1.6$ million unknowns, the point at which the factor still fits one node. Out-of-core factorization and distributed memory are out of our present scope; the next layers for problems that outgrow a node, which we leave to future
work~\cite{ren2025caracal,ren2025ooc}. A GPU is a third such layer, and Section~\ref{sec:gpu} reports a first measurement of the dense route on one device rather than a treatment of it. No solver is best at every scale. What we contribute is one that spans the whole sparse-to-dense spectrum well on shared memory.

\paragraph{Outline.} Section~\ref{sec:patterns} presents the $60$-matrix
benchmark suite and the INLA workload it must serve.
Section~\ref{sec:ordering} describes the fill-reducing ordering and
symbolic factorization that precede every route. Section~\ref{sec:format} formalizes the active-column tile, and
Section~\ref{sec:symbolic} covers the factorization routes: the per-matrix
selector and the fixed-grid (semisparse/dense) and supernodal (sparse) algorithms. Section~\ref{sec:experiments} presents the experimental study, closing with a first extension of the dense route to a GPU (Section~\ref{sec:gpu}). Section~\ref{sec:related} positions the design against prior sparse direct, tile, low-rank, and auto-tuning work, and Section~\ref{sec:conclusion} concludes; two appendices tabulate the per-matrix factorization times on the Intel and AMD nodes, and a third measures the storage of the active-column tile.

\section{The Structural Spectrum and the INLA Workload}\label{sec:patterns}

\begin{figure}[!ht]
  \centering
  \captionsetup[subfigure]{font=small,skip=2pt}
  \begin{subfigure}[t]{0.24\textwidth}
    \centering
    \includegraphics[width=\linewidth]{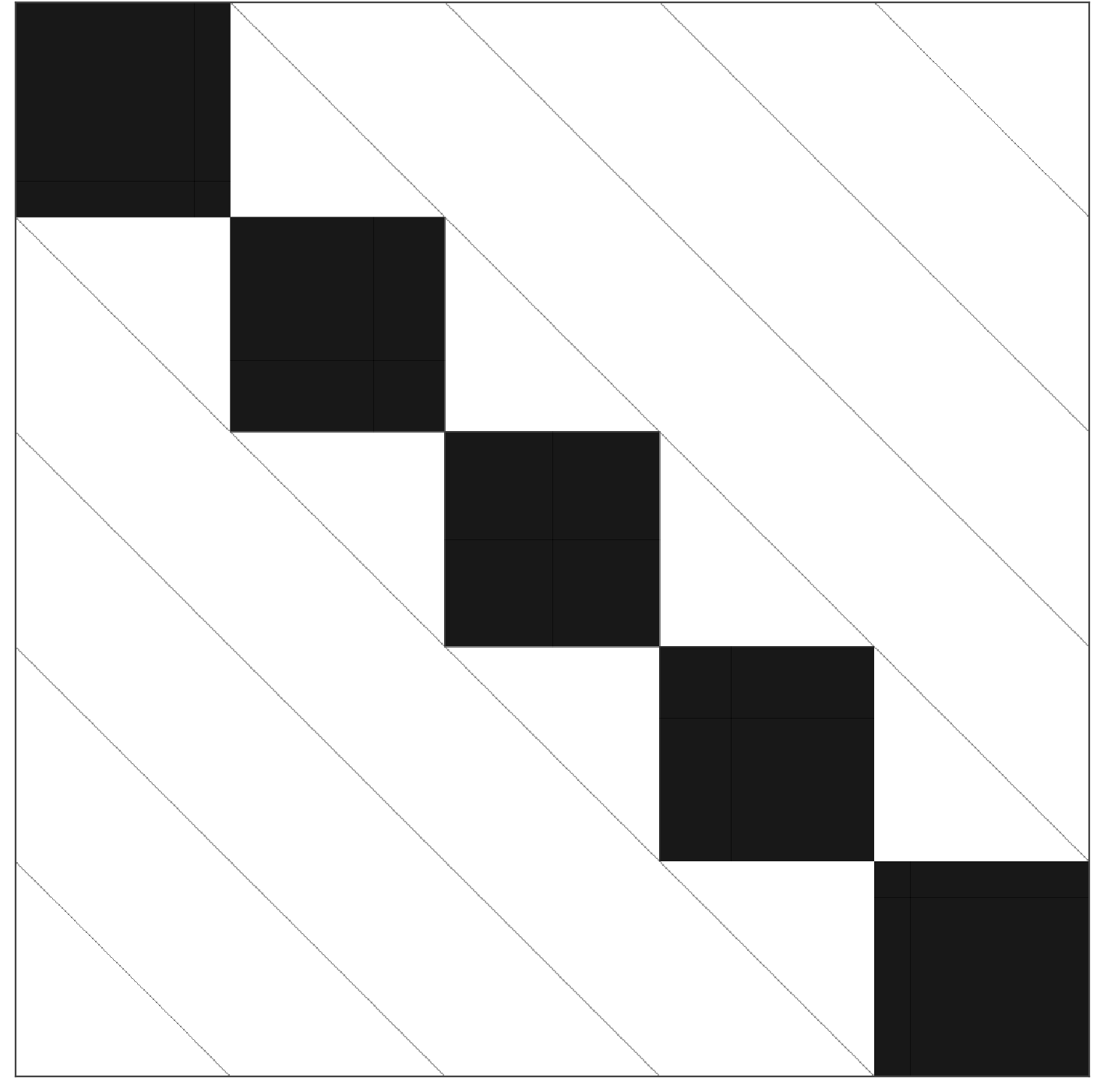}
  \end{subfigure}\hfill
  \begin{subfigure}[t]{0.24\textwidth}
    \centering
    \includegraphics[width=\linewidth]{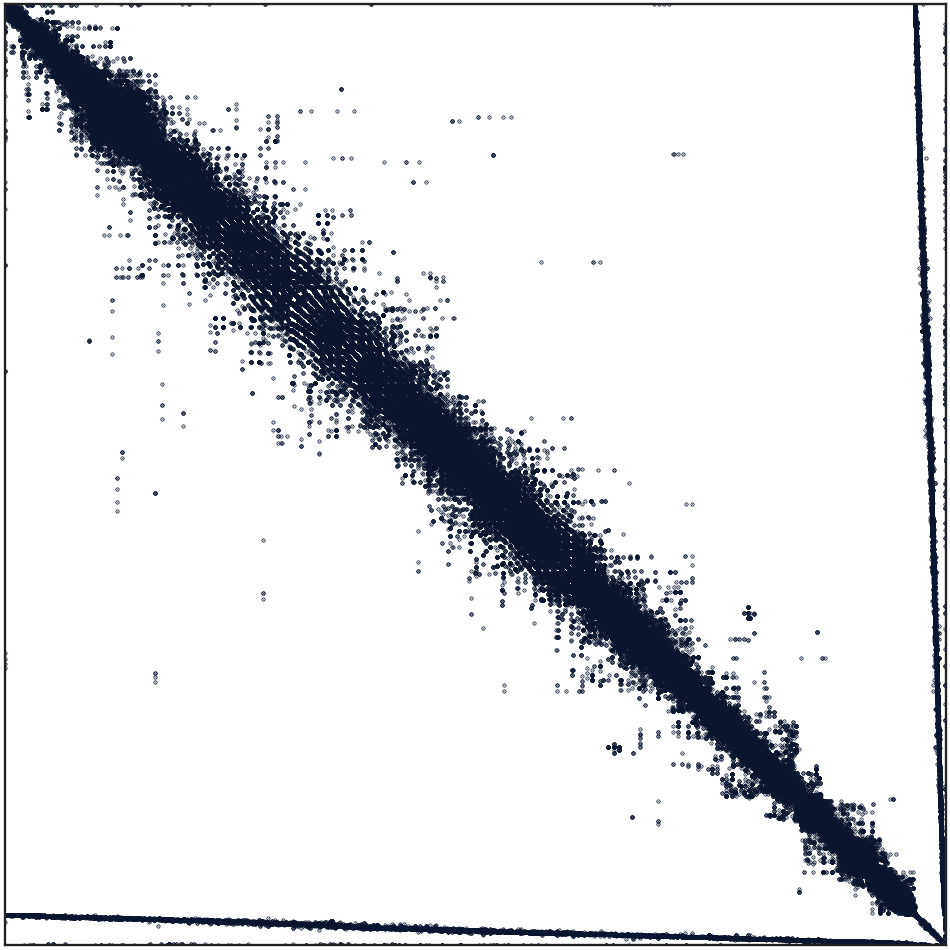}
  \end{subfigure}\hfill
  \begin{subfigure}[t]{0.24\textwidth}
    \centering
    \includegraphics[width=\linewidth]{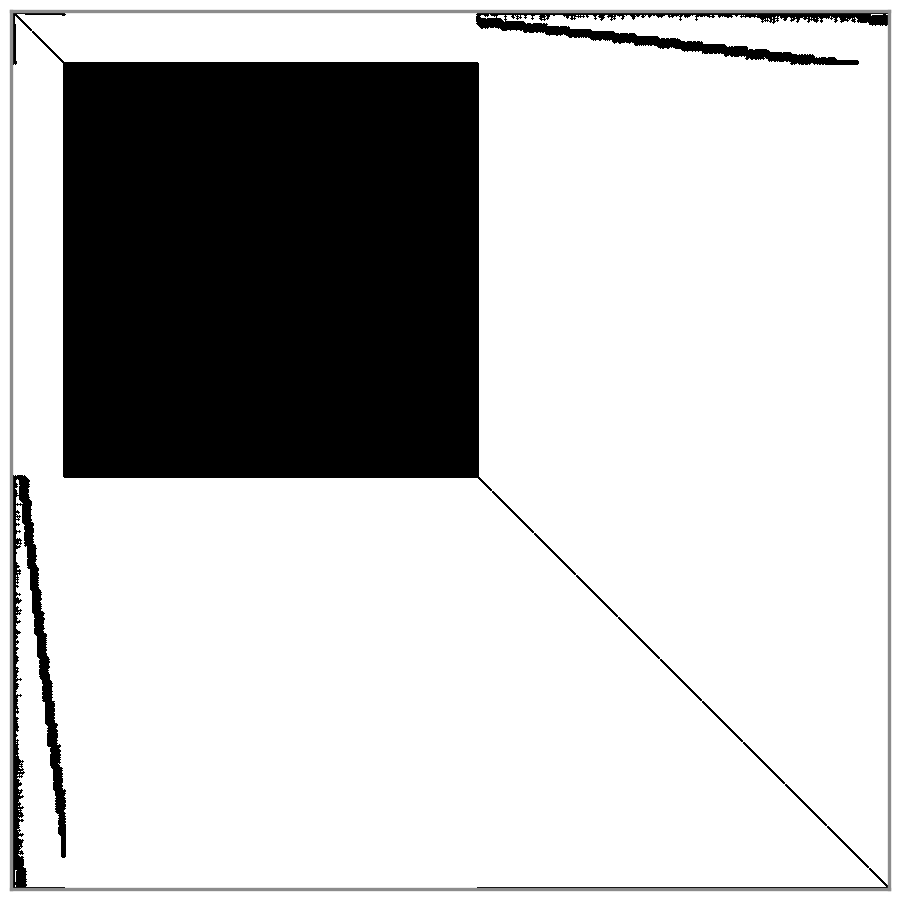}
  \end{subfigure}\hfill
  \begin{subfigure}[t]{0.24\textwidth}
    \centering
    \includegraphics[width=\linewidth]{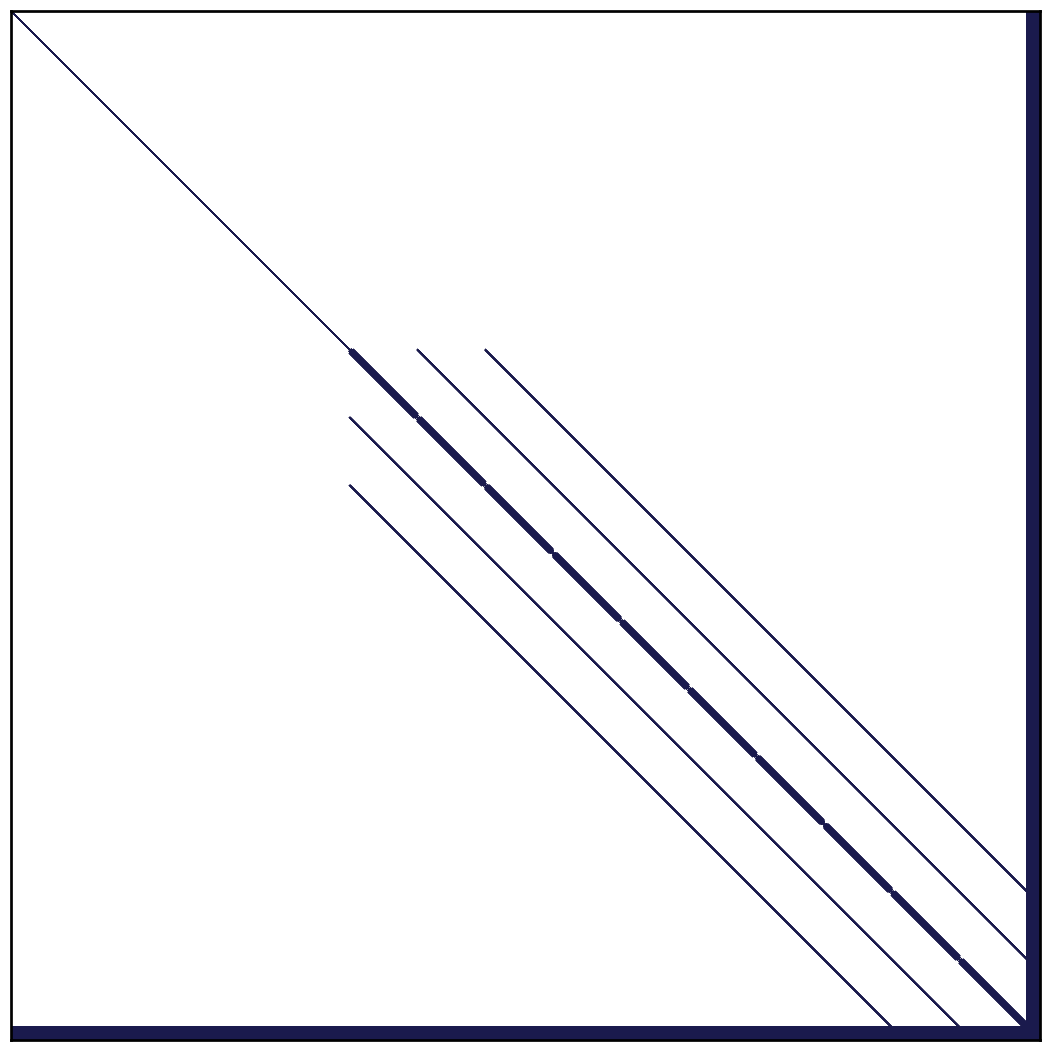}
  \end{subfigure}
    \caption{Representative sparsity patterns from the benchmark suite. From
  left to right: (a) a pedigree-based animal model from quantitative
  genetics, equal dense diagonal blocks coupled by sparse bands; (b) a tight
  band from a PDE discretization; (c) a latent-Gaussian
  precision dominated by one dense block against a sparse remainder; (d) a
  spatiotemporal precision combining a diagonal block, a space-time band,
  and a dense covariate border. Panel (b) is uniformly sparse and routes to
  the supernodal path; the other three mix dense and sparse structure within
  a single matrix, the heterogeneity that motivates a per-tile
  representation.}

  \label{fig:patterns}
\end{figure}

The introduction put concrete numbers on the structural range a single
solver must cover. This section gives the evidence behind those numbers and
describes the workload that produces them. Our suite holds $60$ sparse SPD
matrices from two complementary sources: $26$ latent-Gaussian-model precision
graphs (each the inverse of a covariance) that arise in INLA-based Bayesian
inference~\cite{rue2009approximate}, and $34$ matrices from the SuiteSparse
Matrix Collection (formerly the University of Florida Sparse Matrix
Collection)~\cite{davis2011florida}, drawn from structural and mechanical engineering, thermal, mesh, and
structured-grid applications. The
$26$ INLA graphs are the target workload for which the solver is built; the $34$
SuiteSparse matrices constitute an independent, method-agnostic control, chosen
without reference to how sTiles routes them. We report every matrix and break
results down by regime, so the sparse-regime lead can be traced to the
standard engineering matrices rather than the INLA graphs. Between them the
two sources cover the full structural spectrum the solver must handle
(Table~\ref{tab:suite}).

\begin{table}[t]
\centering\small
\caption{Characteristics of the $60$-matrix benchmark suite by source
  class: order $n$, non-zeros per row $\mathrm{nnz}(A)/n$ and density
  $\mathrm{nnz}(A)/n^2$, both on the full symmetric pattern including the
  diagonal, and fill ratio $\mathrm{nnz}(L)/\mathrm{nnz}(\mathrm{tril}\,A)$,
  which is $1$ for a dense matrix. Each cell gives the range spanned
  across the matrices in that class.}
\label{tab:suite}
\begin{tabular}{@{}lccccc@{}}
\hline
\textbf{Source class} & \textbf{\#} & \textbf{Order $n$} &
  \textbf{nnz/row} & \textbf{Density (\%)} & \textbf{Fill} \\
\hline
INLA Bayesian graphs    & $26$ & $5.1\text{k}$--$1.6\text{M}$ & $3.6$--$10{,}031$ & $7{\times}10^{-4}$--$22$  & $1.0$--$28\times$ \\
SuiteSparse (finite-element method, FEM, etc.) & $34$ & $3.9\text{k}$--$1.0\text{M}$ & $5.0$--$395$      & $5{\times}10^{-4}$--$4.0$ & $2.3$--$78\times$ \\
\hline
Full suite              & $60$ & $3.9\text{k}$--$1.6\text{M}$ & $3.6$--$10{,}031$ & $5{\times}10^{-4}$--$22$  & $1.0$--$78\times$ \\
\hline
\end{tabular}
\end{table}

Table~\ref{tab:suite} summarizes the suite. It spans more than two orders
of magnitude in order $n$, more than three in non-zeros per row (from $3.6$
to over $10{,}000$), more than four in density, and from near-unity to nearly
$80\times$ in fill ratio. The two source classes sit in different corners of
this space. The INLA graphs range from extremely sparse trees to near-dense
precision blocks but mostly incur modest fill, whereas the SuiteSparse
finite-element and structural matrices have moderate per-row density yet
generate the heaviest fill under elimination. No single supernodal
granularity fits this range. A solver tuned for the sparse end, with small
supernodes and fine-grained tasks, underuses BLAS-3 on the dense and
clustered matrices, while one tuned for the dense end, with large supernodes
and coarse tasks, produces excessive fill on the topology-driven graphs. The
same tension appears \emph{within} a single matrix whenever dense blocks sit
beside sparse coupling, as they do in a finite-element front, where no global
granularity resolves it. The per-tile adaptation of
Section~\ref{sec:format} is our answer to it.

Integrated nested Laplace approximation
(INLA)~\cite{rue2009approximate} exercises this entire range
on its own.
It is one framework, but
the latent Gaussian models its users build give rise to structurally distinct
precision patterns. An areal disease-mapping
model~\cite{schrodle2011spatiotemporal,abdulfattah2022interaction} yields a graph that follows an adjacency structure. A spatial or spatio-temporal model discretized through a stochastic partial differential equation
(SPDE)~\cite{lindgren2011explicit} yields a banded or block-banded
field~\cite{cameletti2013spatiotemporal,illian2013fitting}. A genetic or
pedigree animal model yields hierarchical dense blocks coupled by sparse
cross-terms, and a smoothing or temporal-trend component yields a
near-tridiagonal band. Two analysts who run the same software on different
data thus hand the solver different patterns, and even one analyst rarely
knows which pattern a chosen model will produce until the precision is
assembled. A solver tuned in advance to one shape, as the original
arrowhead-only formulation was, therefore serves only the users who happen to build that shape. The range also recurs \emph{within} a single fit. A spatial or temporal field is an extremely sparse graph, while the fixed-effect and covariate terms couple to the whole field and form the dense border that gives the precision its arrowhead structure, with densely connected components filling in further. (Linear constraints add a dense step of their own. INLA enforces them by conditioning by kriging~\cite{rue2009approximate}, which forms and factors a fully dense matrix in the number of constraints around the sparse solve, rather than densifying the precision the solver factors.) A solver for INLA must therefore handle patterns from graph-sparse to a
sparse field carrying a few fully dense rows and columns, both across users
and within one fit.

\section{Ordering and Symbolic Factorization}\label{sec:ordering}

The per-tile representation of Section~\ref{sec:format} and the per-matrix
routing of Section~\ref{sec:symbolic} both draw on a single preprocessing
pass, run once per sparsity pattern and reused across every factorization of
it. It begins by building the adjacency graph of $A$ at the \emph{element}
level, with effectively dense rows set aside so that they do not distort the
ordering heuristics. Rather than commit to one fill-reducing heuristic on
that graph, sTiles runs a small set of them in parallel, including reverse
Cuthill--McKee (RCM)~\cite{cuthill1969reducing}, approximate minimum degree
variants~\cite{amestoy2004amd,davis2004colamd},
METIS~\cite{karypis1998metis}, and SCOTCH~\cite{chevalier2008pt}.

Each candidate permutation is then scored, and the scoring is what makes the
pass expensive: the quality of a permutation is not predictable in advance,
so every candidate is put through a full symbolic Cholesky factorization of
its own. This is the step Figures~\ref{fig:tilemodes-a}
and~\ref{fig:tilemodes-b} illustrate, turning the sparsity pattern of $A$
into the filled factor $L$, and it yields two numbers per candidate. The
first is the fill $\texttt{nnz}(L)$. The second is obtained by mapping the
pattern onto a uniform grid of tiles and counting how many tiles it makes
active, and it is the one that matters most for a tile solver, because the
tile is the unit of work: an active tile is processed by whichever kernel its
route later selects, and an empty tile is skipped altogether. Two
permutations with the same fill can therefore cost very different amounts.
Fill that falls within what a tile already holds, whether a dense block or a
set of active columns, raises the average number of elements per tile and is
largely absorbed, since that memory is traversed as a block in any case,
whereas fill that falls outside it widens a tile with further active columns
or activates tiles that were empty. Extra fill can consequently be an
advantage rather than a penalty.

The winner is chosen accordingly. On graphs sparse enough that most tiles
stay near-empty, $\texttt{nnz}(L)$ is a fair proxy for the work and sTiles
selects on it; on denser graphs it selects on the number of active tiles
instead, preferring a permutation that packs the fill more tightly even when
its total fill is higher, and setting that preference aside only for a
candidate whose fill is lower by enough to outweigh the extra tiles, the
arithmetic growing faster than linearly in $\texttt{nnz}(L)$. A separate
allowance admits nested-dissection orderings at a bounded fill penalty on
large problems, where the shallow separator tree is what exposes any
parallelism at all. The pattern computed for the winning candidate is kept
and those of the others are discarded, which is where most of the cost of the
pass goes.

Three features are then read off the winning pattern on the same tile grid,
the mean tile occupancy, the fill ratio, and the degree skew, and the matrix
is routed on them (Section~\ref{sec:selector}). Only after that does the
layout diverge: the dense and semisparse routes keep the uniform grid, while
the sparse route rebuilds the pattern in CSC form and runs a supernodal
symbolic factorization that amalgamates columns into supernodes of unequal
width. Allocating the factor and building the static task graph follow, both
minor in cost beside the search.

Running several orderings makes the analysis more expensive than committing
to one, but the trade is deliberate: it is paid once per pattern. We add that the ordering is not what drives the performance reported in
Section~\ref{sec:experiments}; it yields a factor comparable in size to that of a
conventional supernodal solver rather than a smaller one, and on several
matrices sTiles is faster while forming more fill.

\section{The Active-Column Tile}\label{sec:format}

The active-column tile is our answer to the regular middle of the structural
range; Section~\ref{sec:grid-alignment} identifies the structures it suits. We write the factorization $A = LL^\top$ with $L$ lower triangular and use $L$ throughout. The implementation stores the transpose $L^\top$, a layout choice with no effect on the mathematics.

\begin{figure*}[t]
  \centering
  \captionsetup[subfigure]{font=small,justification=centering,
                           singlelinecheck=true,skip=2pt}
  \begin{subfigure}[t]{0.235\linewidth}
    \centering
    \includegraphics[width=\linewidth]{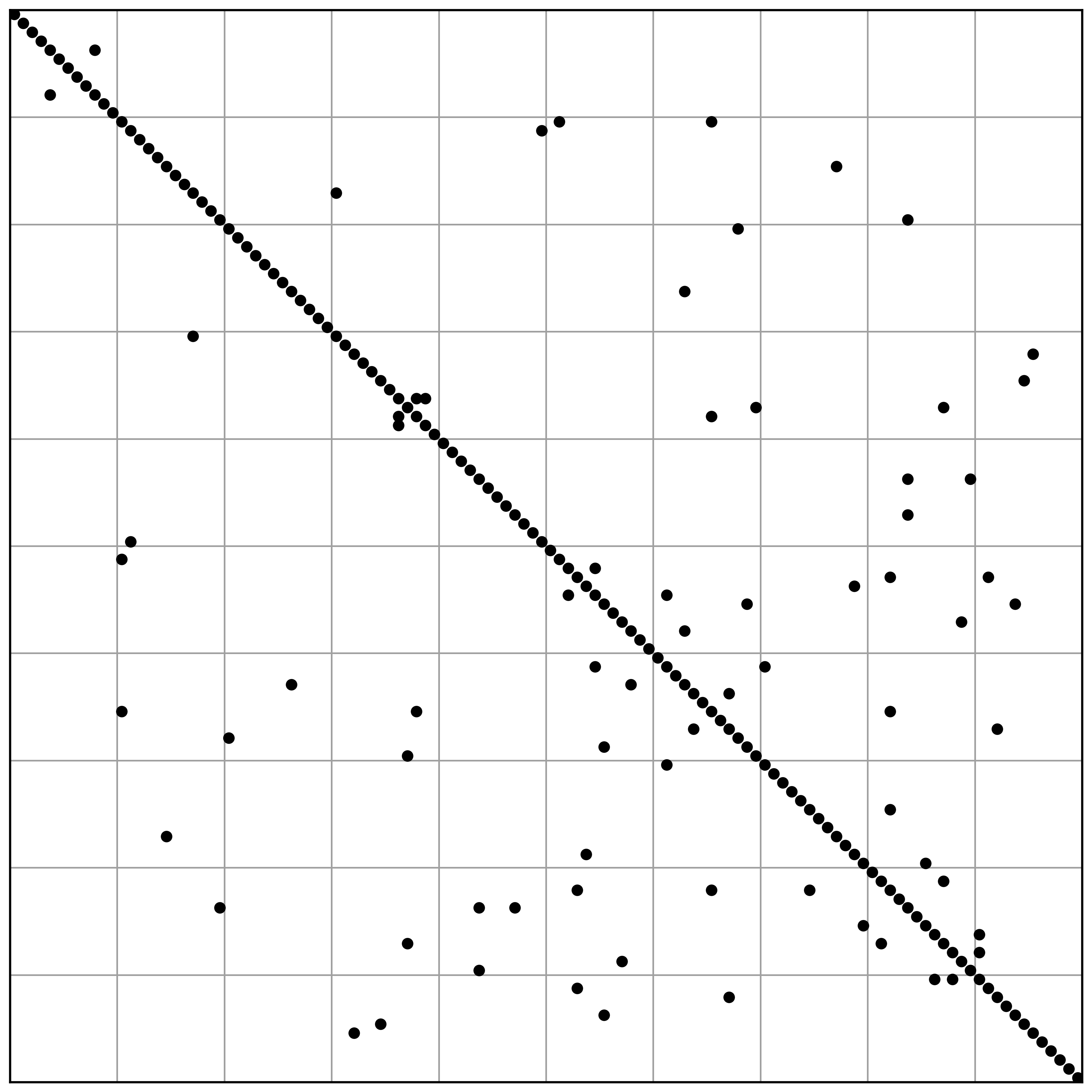}
    \caption{Pattern of $A$.}
    \label{fig:tilemodes-a}
  \end{subfigure}\hfill
  \begin{subfigure}[t]{0.235\linewidth}
    \centering
    \includegraphics[width=\linewidth]{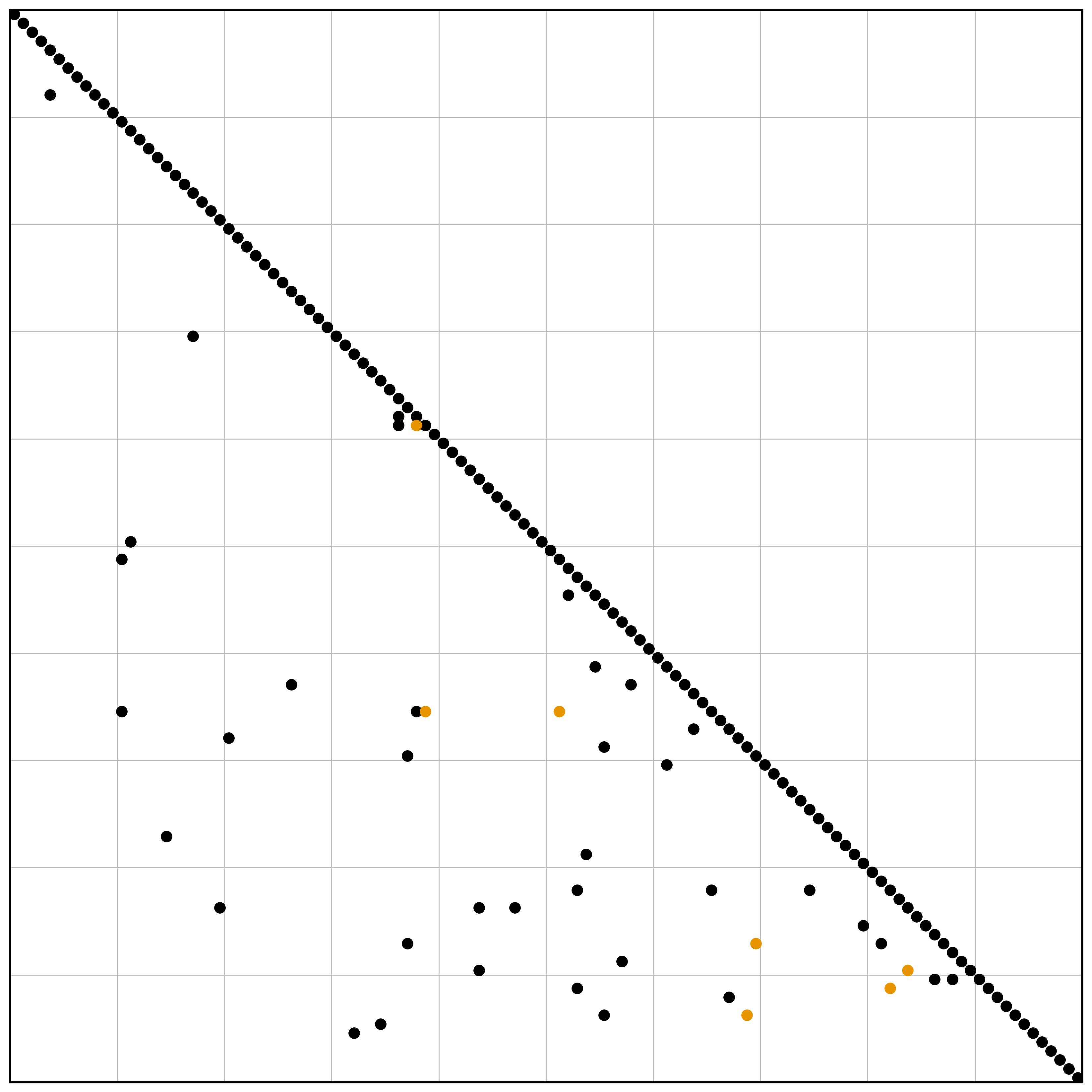}
    \caption{Factor $L$ with fill-in.}
    \label{fig:tilemodes-b}
  \end{subfigure}\hfill
  \begin{subfigure}[t]{0.235\linewidth}
    \centering
    \includegraphics[width=\linewidth]{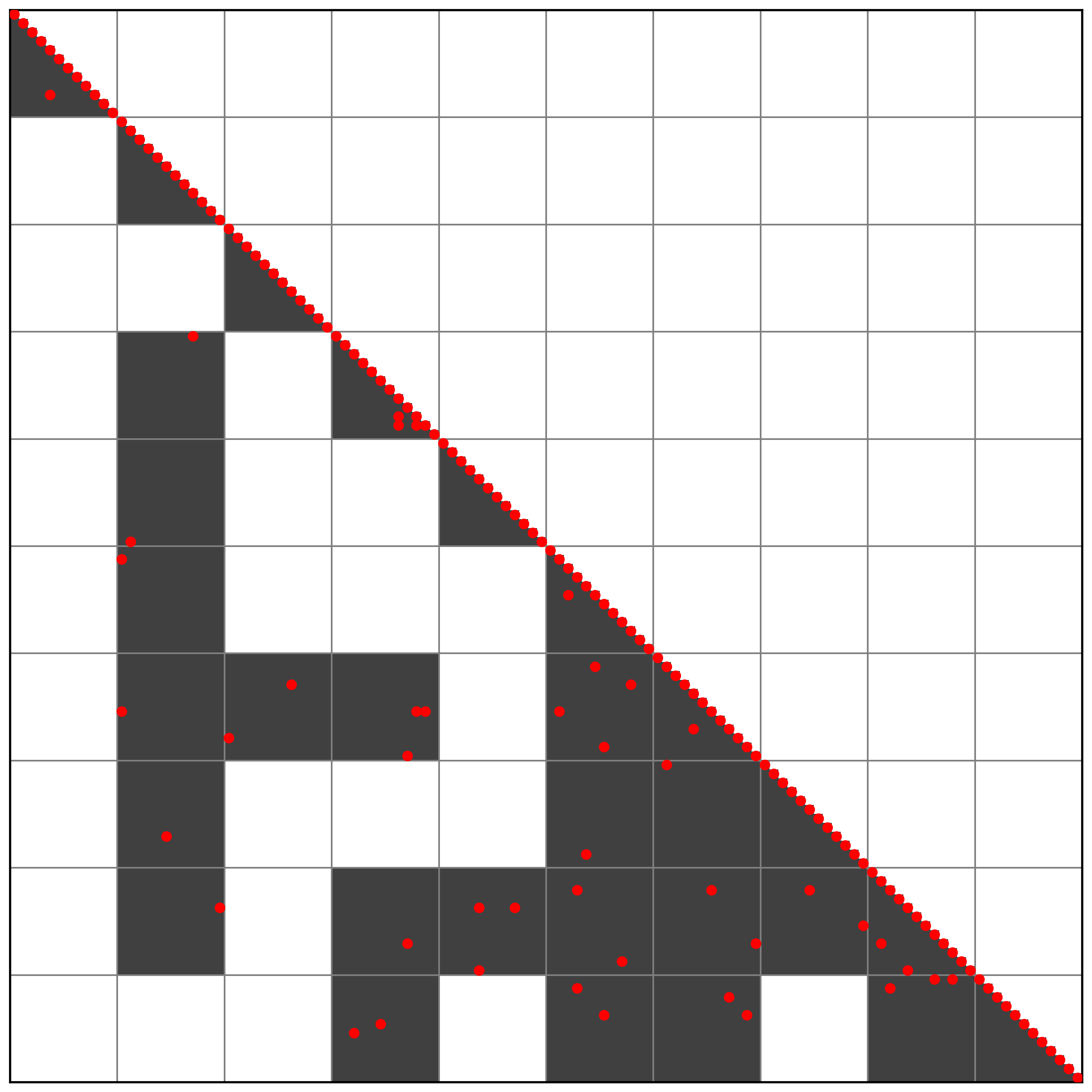}
    \caption{Dense tiling.}
    \label{fig:tilemodes-c}
  \end{subfigure}\hfill
  \begin{subfigure}[t]{0.235\linewidth}
    \centering
    \includegraphics[width=\linewidth]{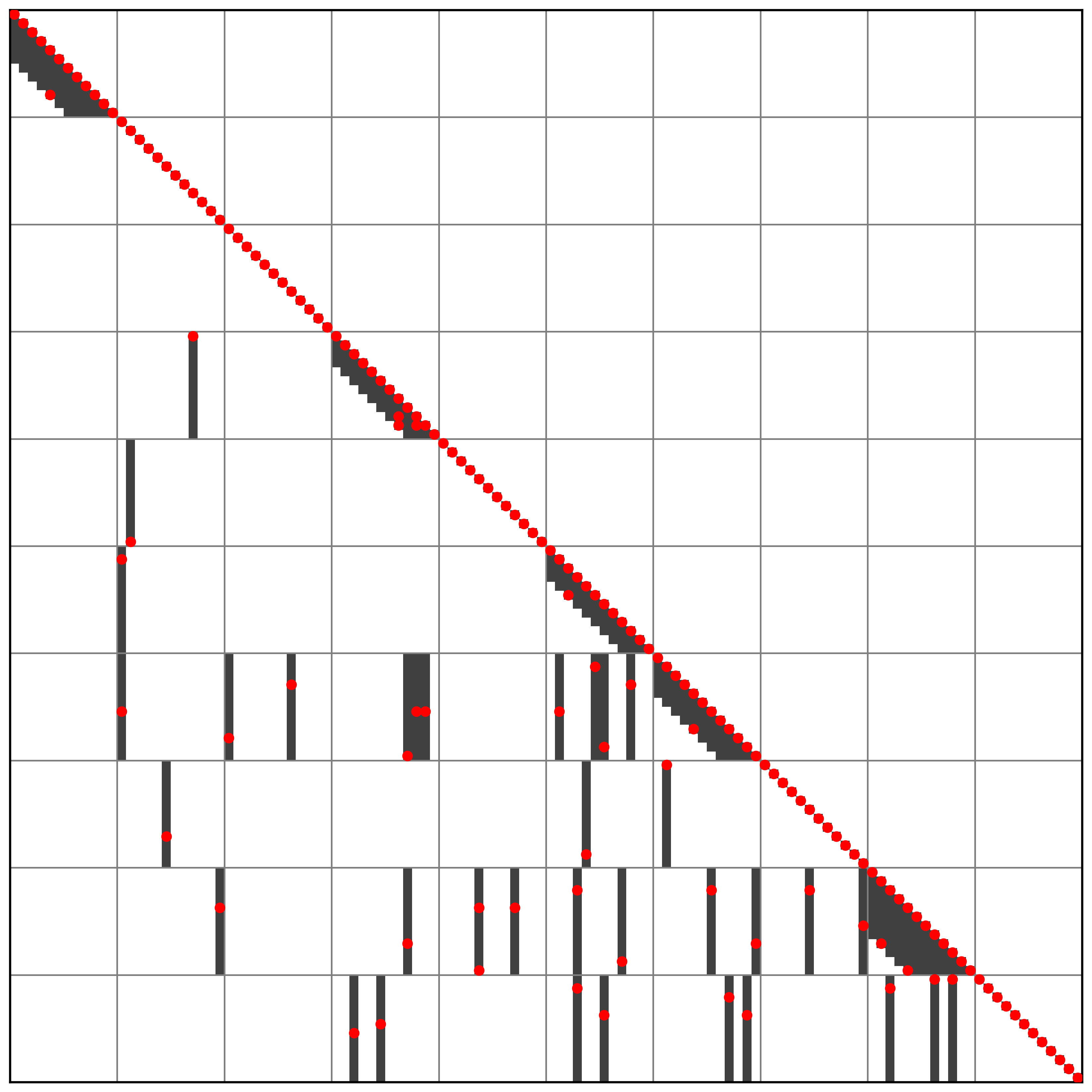}
    \caption{Active-column tiling.}
    \label{fig:tilemodes-d}
  \end{subfigure}
  \caption{From sparsity pattern to tile storage on a random
    symmetric SPD instance ($N = 120$, $n_b = 12$).
    (\subref{fig:tilemodes-a})~the non-zero pattern of $A$;
    (\subref{fig:tilemodes-b})~the symbolic Cholesky factor $L$, with
    original $A$ entries in black and fill-in entries in orange;
    (\subref{fig:tilemodes-c})~dense tile storage allocates a full
    $n_b \times n_b$ buffer for every active off-diagonal tile and
    the upper-triangular half of every active diagonal tile (the
    lower triangle of a diagonal tile is zero by symmetry and is
    not stored);
    (\subref{fig:tilemodes-d})~active-column tile storage holds each diagonal tile, of half-width \texttt{kd}, in $(\texttt{kd}+1) \times n_b$ banded-packed
    form and each off-diagonal tile in $n_b \times |\mathcal{A}|$
    active-column form, dropping both the inactive columns and the
    band's upper-triangular zeros. On this instance, the active-column
    layout stores roughly $22\%$ of the entries that the dense
    layout does; the savings widen as $n_b$ grows relative to the
    local fill density.}
  \label{fig:tilemodes}
\end{figure*}

Throughout, a sparse SPD matrix of order $N$ is partitioned into an $N_T
\times N_T$ grid of $n_b \times n_b$ tiles, with $N_T = \lceil N / n_b
\rceil$. The grid is fixed before any numerical work and is the same across all modes; only the storage of each tile changes. Any tile the symbolic factorization shows to be structurally empty is dropped entirely, neither stored nor scheduled, so that only occupied tiles enter the factorization and ``tiling a sparse matrix'' never materializes the zeros between tiles. A \emph{dense tiling} stores each occupied tile as a full $n_b \times n_b$ block and carries the zeros enclosed \emph{within} that tile. The \emph{active-column tiling} we develop here keeps only the tile's active columns and drops the zeros in its inactive ones. (The active columns are still stored at full height, zeros included, so that a dense BLAS-3 kernel can run on them.) Figure~\ref{fig:tilemodes} shows the contrast that motivates this section at the level of a whole sparse SPD matrix: the same Cholesky factor under a full dense tiling (Fig.~\ref{fig:tilemodes-c}) and under the active-column tiling (Fig.~\ref{fig:tilemodes-d}). The compression itself works
one tile at a time. Consider a single $n_b \times n_b$ tile in the
off-diagonal of a larger factorization (drawn small in the examples below for legibility). Some columns of the tile carry non-zero entries, and others do not, and the symbolic factorization of the matrix settles that distinction once and for all. We call a column \emph{active} if it contains at least one non-zero entry that the numerical factorization will eventually read or write, and write $\mathcal{A} \subseteq \{0, 1, \dots, n_b - 1\}$ for the set of active column indices. When the tile sits inside a sparse region of the matrix, $|\mathcal{A}|$ can be a small fraction of $n_b$; when it sits inside a dense block, $\mathcal{A}$ is all of $\{0, \dots, n_b - 1\}$.

\subsection{Active columns and compression}\label{sec:active-cols}

A tile taken from the matrix is a region of $n_b \times n_b$ entries, most of them zero, and storing all of them densely would waste space on columns the symbolic phase has already found to be empty. Compression works at the column level rather than the entry level, because a column either carries non-zeros, in which case the kernel must touch it, or it does not, in which case the kernel can skip it entirely. It proceeds in three stages, shown below for a tile with active set
$\mathcal{A} = \{0, 2, 3\}$:
\[
  \underbrace{\begin{bmatrix}
    0 & 0 & \bullet & 0 & 0 \\
    \bullet & 0 & 0 & \bullet & 0 \\
    0 & 0 & \bullet & 0 & 0 \\
    \bullet & 0 & 0 & \bullet & 0 \\
    0 & 0 & \bullet & \bullet & 0
  \end{bmatrix}}_{\text{(a) elements}}
  \;\to\;
  \underbrace{\begin{bmatrix}
    \bullet & 0 & \bullet & \bullet & 0 \\
    \bullet & 0 & \bullet & \bullet & 0 \\
    \bullet & 0 & \bullet & \bullet & 0 \\
    \bullet & 0 & \bullet & \bullet & 0 \\
    \bullet & 0 & \bullet & \bullet & 0
  \end{bmatrix}}_{\text{(b) active columns}}
  \;\to\;
  \underbrace{\begin{bmatrix}
    \bullet & \bullet & \bullet \\
    \bullet & \bullet & \bullet \\
    \bullet & \bullet & \bullet \\
    \bullet & \bullet & \bullet \\
    \bullet & \bullet & \bullet
  \end{bmatrix}}_{\text{(c) stored}},
  \quad
  \texttt{aind} = (0, 2, 3).
\]
Panel (a) is the original tile after symbolic factorization, where $\bullet$ marks an entry the numerical phase will read or write and $0$ marks one guaranteed to stay zero. Panel (b) is column activation: columns $0$, $2$, and $3$ each hold at least one $\bullet$, so the whole column is flagged \emph{active} and given dense storage, while the empty columns $1$ and $4$ are dropped. Panel (c) is the physical layout, in which only the active columns occupy memory, packed contiguously into an $n_b \times |\mathcal{A}|$ dense buffer, with the index array $\texttt{aind} = (0, 2, 3)$ recording their original positions inside the tile.

The metadata a kernel needs is a small record: the number of active columns, their identities (the index array $\texttt{aind}$), and a flag for the common full-width fast case $\mathcal{A} = \{0, \dots, n_b - 1\}$. The stored tile is itself a small dense matrix of shape $n_b \times |\mathcal{A}|$, so BLAS-3 runs on it unchanged; the format alters what the BLAS sees, not how it computes. It is not a full dense tile, which would pay for the inactive columns, nor a sparse storage format, which would forfeit BLAS-3 efficiency: it drops the columns the symbolic phase has proven empty and keeps the rest dense. The one size-dependent exception is that when a tile's active dimension is small (at most a threshold, $16$ by default), the rank-$k$ and matrix-multiply updates run through a hand-rolled, SIMD-vectorized kernel that fuses the multiply with the scatter, because at that size the dispatch overhead of a BLAS-3 call exceeds the arithmetic it performs.

\subsection{Scatter and gather}\label{sec:scatter}

Compression comes at a cost: a kernel that reads from or writes to the tile must translate between two coordinate systems. The \emph{logical} column index $j \in \{0, \dots, n_b - 1\}$ is the column's position inside the full $n_b \times n_b$ tile, and hence inside the matrix the tile belongs to. The \emph{stored} column index $j' \in \{0, \dots, |\mathcal{A}| - 1\}$ is its position in the packed dense buffer. The forward map $j' \mapsto j$ is just $\texttt{aind}[j']$; the inverse map $j \mapsto j'$ is built on demand and is usually consumed implicitly through a precomputed \emph{scatter} array.

A kernel whose result is indexed in the stored coordinate system of an input tile, but which must be written into the logical coordinate system of an output tile, performs a \emph{scatter}: column $j'$ of the result goes to column $\texttt{aind}_{\text{out}}[j']$ of the output. With $\texttt{aind} = (0,2,3)$, stored columns $0,1,2$ land in logical columns $0,2,3$ and logical columns $1,4$ are left unwritten. The executor precomputes this column map once per task, so the kernel's inner loop reads a single integer per output column instead of walking a comparison loop. Which case applies is settled during preprocessing rather than tested at runtime: the symbolic phase records with each tile whether its active set is contiguous and whether it is full width. A contiguous active set collapses the scatter to a base offset, letting the kernel accumulate into consecutive entries with no indirection, and a full-width set ($|\mathcal{A}| = n_b$) removes the scatter altogether.

The reverse operation, \emph{gather}, reads a logical column $j$ from a tile by looking $j$ up in $\texttt{aind}$ to get the stored position $j'$ and reading that column straight from the packed buffer. Since the stored buffer is already a column-major dense matrix, each gather is one pointer offset per column read.

\subsection{POTRF on a diagonal tile}\label{sec:potrf}

Diagonal tiles are stored differently from off-diagonal ones. Because the
matrix is SPD, its diagonal is always non-zero, so the active set
$\mathcal{A}$ of a diagonal tile is the whole range $\{0, \dots, n_b - 1\}$ and the active-column compression of Section~\ref{sec:active-cols} becomes a no-op. What is non-trivial about a diagonal tile is its
\emph{bandwidth}. Factorization within the tile fills a band of half-width
$\texttt{upper\_bw} = \texttt{kd}$ around the diagonal, and the tile is stored
physically in LAPACK's~\cite{anderson1999lapack} upper-banded packed format.
For a tile with $\texttt{kd} = 2$,
\[
  \underbrace{\begin{bmatrix}
    a_{00} & a_{01} & a_{02} & 0      & 0      \\
    a_{01} & a_{11} & a_{12} & a_{13} & 0      \\
    a_{02} & a_{12} & a_{22} & a_{23} & a_{24} \\
    0      & a_{13} & a_{23} & a_{33} & a_{34} \\
    0      & 0      & a_{24} & a_{34} & a_{44}
  \end{bmatrix}}_{\text{banded, } \texttt{kd} = 2}
  \;\longrightarrow\;
  \underbrace{\begin{bmatrix}
    \ast   & \ast   & a_{02} & a_{13} & a_{24} \\
    \ast   & a_{01} & a_{12} & a_{23} & a_{34} \\
    a_{00} & a_{11} & a_{22} & a_{33} & a_{44}
  \end{bmatrix}}_{\text{packed, } (\texttt{kd} + 1) \times n_b}.
\]
The packed layout drops the zero corners of the band and keeps only the main
diagonal and the $\texttt{kd}$ super-diagonals, in $(\texttt{kd} + 1) \times n_b$
column-major order: entry $a_{ij}$ with $i \le j$ and $j - i \le \texttt{kd}$ lives at storage position $(\texttt{kd} + i - j, j)$, the main diagonal occupies the bottom row of the buffer, and the entries marked $\ast$ in the top-left corner go unused. This costs $(\texttt{kd} + 1) \, n_b$ doubles, asymptotically smaller than the $\tfrac{1}{2} n_b^2$ a dense upper-triangular block would need when $\texttt{kd} \ll n_b$.

The \textsc{potrf} kernel for a diagonal tile is then a single banded LAPACK call. The \textsc{syrk} updates that accumulate into $L_{kk}$ are earlier tasks in the same thread's slice of the static schedule, since the panel partition places every write to a column on one thread (Section~\ref{sec:executor}), so they are already applied by the time this task runs. The worker thread then invokes the banded LAPACK factorization \textsc{dpbtrf} directly on the packed buffer:
\[
  L_{kk} \;\longleftarrow\; \textsc{dpbtrf}\bigl(L_{kk},\, \texttt{kd}\bigr).
\]
No scatter or gather is involved. The factor overwrites the banded buffer in place, the kernel marks the tile ready through the executor's progress matrix, and the downstream \textsc{trsm} tasks for the same column start executing. The bandwidth $\texttt{kd}$ is set once by the symbolic phase, from the largest off-diagonal fill the tile receives during factorization, and the tile's storage is allocated to match, so no resize is ever needed at numerical time.

The off-diagonal kernels (\textsc{trsm},~\textsc{syrk},~\textsc{gemm}) follow the same compression
discipline but consume two or three tiles whose active sets need not agree. The executor precomputes a scatter for each input-to-output pair, so each kernel runs as a BLAS-3 call on the packed sub-blocks and scatters its result into the destination tile's active columns (Section~\ref{sec:scatter}).

\subsection{Related storage formats}\label{sec:related-storage}

The storage primitives the active-column tile\footnote{We use \emph{active-column tile} for this per-tile storage format and reserve \emph{semisparse} for the class of matrices it serves; the latter is unrelated to the semi-separable matrices and the semi-sparse formats of other subfields, which share only the name.} builds on are not new. Column masking, packed dense sub-blocks, and zero-padded supernodes all predate it. What it adds is narrow and lies in the combination: intra-tile column masking inside a direct factorization on a fixed, geometry-driven grid, in which the dense tile is the literal fully-active limit of one code path. The payoff is in the execution model rather than in storage. The same dense BLAS-3 kernels are reached on a partition fixed ahead of time, so the schedule amortizes across refactorizations (Section~\ref{sec:related}). Here we note only how the storage itself differs from established formats, in three respects.

First, inactive columns take no storage and are touched by no kernel. In an off-diagonal tile the stored width is the \emph{exact} active-column count, so a near-empty tile costs only its metadata plus that count's worth of full-height arithmetic, while diagonal tiles instead drop their band's zeros (Section~\ref{sec:potrf}). Either way the width is never inflated to a worst case. Against an exact sparse format, the residual overhead is the zeros inside selected active columns; measured on the matrices the selector routes here it is $1.22\times$ the nonzeros of $L$, but $0.81\times$ its memory, since the tile carries no per-entry row index (Appendix~\ref{app:storage}). This contrasts with sliced ELLPACK and
SELL-C-$\sigma$~\cite{monakov2010ellpack,kreutzer2014unified}, which set a
padding width per row-slice from that slice's longest row, independent of which columns carry active data; the active-column tile instead drops inactive columns outright, so its width tracks column activity rather than row length. Second, masking columns \emph{out} is the dual of supernode relaxation. Amalgamated and relaxed supernodes~\cite{ashcraft1989relaxed} add explicit zeros to merge thin
columns into dense blocks for BLAS efficiency, whereas the active-column tile
starts from a dense block and strips the inactive columns, reaching dense BLAS-3 from the opposite direction. Third, the discipline is applied within a \emph{direct factorization} at intra-tile granularity, where the prevailing approach is the reverse: let the matrix dictate the partition and absorb the resulting task-size irregularity in the
runtime~\cite{amestoy2001fully,henon2002pastix}.


\subsection{When the tile pays: grid alignment}\label{sec:grid-alignment}
The three fast paths above are not generic, and taken together they pin down which structures the active-column tile suits. A fully active tile costs a dense block plus one accumulate pass; its scatter carries no indirection (Section~\ref{sec:scatter}). A tile whose active columns are \emph{contiguous} costs a single pointer offset instead of an indexed scatter. A diagonal tile costs $(\texttt{kd}+1)\,n_b$ rather than $\tfrac12 n_b^2$ only when its bandwidth satisfies $\texttt{kd} \ll n_b$ (Section~\ref{sec:potrf}). A \emph{regular band beside a dense border} hits all three at once: the border fills whole tiles (the no-op path), each off-diagonal band tile fills a contiguous run of columns (the pointer-offset path), and the diagonal band stays narrow (the banded path). No kernel walks a scatter map, and the factor is far smaller than a dense one. This is the banded-bordered structure of the latent-Gaussian precisions, whose diagonal half-width holds at a small fraction of the tile. An \emph{irregular} front defeats all three paths at once: its active columns are scattered rather than contiguous, so every update
pays the indexed scatter; its diagonal fills to $\texttt{kd} \approx n_b$, so the banded packing saves nothing; and few tiles are full. There, a fixed grid is the wrong partition, and the supernodal route, whose variable-size tiles follow the fill, wins, which is why the selector sends such matrices to it (Section~\ref{sec:selector}). The active-column tile is therefore a sharp instrument for the regular banded or bordered middle, not a general sparse format. Its breadth in the solver comes from the selector that pairs it with the dense and supernodal routes, not from the tile on its own.

\section{Factorization Routes}\label{sec:symbolic}

Once the matrix is reordered and symbolically factored
(Section~\ref{sec:ordering}), the solver assigns it to one of three routes:
the active-column tile, a plain dense tiling, or a non-uniform supernodal
tiling. The choice comes from the symbolic factor alone, before any numerical
work, and it fixes the data structure, the task graph, and the kernels for
every factorization of that pattern. It is therefore paid once and reused
across the repeated factorizations the design is built for. This section
covers the selector that picks the route (Section~\ref{sec:selector}), the
fixed-grid algorithm behind the semisparse and dense routes
(Section~\ref{sec:tilealg}), the static shared-memory executor that runs the
resulting task list (Section~\ref{sec:executor}), and the supernodal tiling
that serves the sparse route (Section~\ref{sec:supernodal}).

\subsection{Choosing the representation}\label{sec:selector}

From the symbolic factor, computed once per pattern, the selector reads three
inexpensive features: the fill ratio $\phi = \texttt{nnz}(L)/\texttt{nnz}(A)$; the
mean off-diagonal block-density $\rho$, the mean occupancy of the active
off-diagonal tiles of $L$ (occupancy being occupied columns over $n_b$); and
the \emph{degree skew} $\sigma$, the largest column count of $L$ over the mean,
which flags a dense border. It applies them as a fixed precedence:
\begin{enumerate}[leftmargin=1.5em,topsep=2pt,itemsep=1pt,label=(\roman*)]
\item tiles almost entirely full $\Rightarrow$ \emph{dense} tiling
      (Section~\ref{sec:tilealg});
\item high fill, where active-column masking still leaves each tile mostly
      empty $\Rightarrow$ \emph{supernodal} tiling (Section~\ref{sec:supernodal});
\item otherwise \emph{semisparse}, when tiles are reasonably occupied
      \emph{or} the degree skew reveals a dense border; failing both, fall
      through to supernodal.
\end{enumerate}
The active-column tile is thus the working default: its fully active limit is a
plain dense block (Section~\ref{sec:format}), so it already spans the
dense-to-thin range within one factorization, and the two extremes peel off
only when their signals fire. Sending near-full fronts to the dense route
matters because their high fill ratio would otherwise misroute them to the
supernodal tiling, which loses parallel efficiency on such fronts.

\subsection{Semisparse and dense factorization}\label{sec:tilealg}

This section covers two of the three routes, the \emph{dense} route and the
\emph{semisparse} route. The two share a single left-looking tile
factorization and differ only in how each tile is stored and how the kernels
touch it. Both overlay the same uniform $n_b \times n_b$ tile grid on $L$,
compile to the same task list, and run on the same executor. The symbolic
sweep of Algorithm~\ref{alg:left-looking-semisparse} collects the list once,
as an ordered sequence of the four kernel cases (Case~1~\textsc{potrf} on a
diagonal tile; Case~2~\textsc{syrk} and Case~4~\textsc{gemm} for the
left-of-pivot updates; Case~3~\textsc{trsm} for the off-diagonal panel solve),
and the numeric phase then walks the list and executes each task.

Algorithm~\ref{alg:left-looking-semisparse} states the collection precisely.
Writing $\mathcal{N}(k) = \{\, i \neq k : L_{ik} \neq 0 \ \text{or}\ L_{ki}
\neq 0 \,\}$ for the structural tile neighbours of block $k$ (so that
$n \in \mathcal{N}(k)$ with $n < k$ are the block-columns left of the pivot
that update it, and $m \in \mathcal{N}(k)$ with $m > k$ the nonzero tile rows
below it), it appends, for each $k$, the
left-of-pivot updates, then the diagonal factorization, then the off-diagonal
panel solve. 

\begin{algorithm}[t]
\caption{Left-looking tile Cholesky. A single sweep over the block
structure collects the ordered task list of the four kernels,
1~(\textsc{potrf}), 2~(\textsc{syrk}), 3~(\textsc{trsm}), 4~(\textsc{gemm}).
The numeric executor (Algorithm~\ref{alg:exec}) then runs this list.}
\label{alg:left-looking-semisparse}
\small
\begin{algorithmic}[1]
\State \textbf{Input:} tile count $N_T$; symbolic neighbour sets
       $\mathcal{N}(k) = \{\, i \neq k : L_{ik} \neq 0 \ \text{or}\ L_{ki} \neq 0 \,\}$.
\State \textbf{Output:} ordered task list $\mathcal{T}$; each task is a
       $(\text{case},\, m,\, k,\, n)$ tuple, with $-$ marking an unused slot.
\State $\mathcal{T} \gets \langle\;\rangle$
\For{$k \gets 0$ \textbf{to} $N_T-1$} \Comment{left-looking sweep over block columns}
    \ForAll{$n \in \mathcal{N}(k)$ \textbf{with} $n < k$}
        \State append $(2,\, -,\, k,\, n)$ to $\mathcal{T}$
               \Comment{\textsc{syrk}: $L_{kk} \mathrel{-}= L_{kn} L_{kn}^{\top}$}
    \EndFor
    \State append $(1,\, -,\, k,\, -)$ to $\mathcal{T}$
           \Comment{\textsc{potrf}: factor diagonal tile $L_{kk}$}
    \ForAll{$m \in \mathcal{N}(k)$ \textbf{with} $m > k$}
        \ForAll{$n \in \mathcal{N}(k) \cap \mathcal{N}(m)$ \textbf{with} $n < k$}
            \State append $(4,\, m,\, k,\, n)$ to $\mathcal{T}$
                   \Comment{\textsc{gemm}: $L_{mk} \mathrel{-}= L_{mn} L_{kn}^{\top}$}
        \EndFor
        \State append $(3,\, m,\, k,\, -)$ to $\mathcal{T}$
               \Comment{\textsc{trsm}: $L_{mk} \gets L_{mk} L_{kk}^{-\top}$}
    \EndFor
\EndFor
\State \Return $\mathcal{T}$
\end{algorithmic}
\end{algorithm}

\paragraph{Dense route.}
Every tile is a full, dense $n_b \times n_b$ block, and the four cases are the
standard dense LAPACK and BLAS kernels: \textsc{potrf} factors each diagonal
tile, \textsc{trsm} solves each off-diagonal tile against it, and \textsc{syrk}
and \textsc{gemm} apply the left-of-pivot updates to the diagonal and the rest
of the panel. This is the dense tiled Cholesky of the original
sTiles framework~\cite{abdulfattah2025stiles}, extended here with block-level
sparsity: structurally zero tiles are pruned from the task list, so the sweep
factors only the nonzero tiles rather than the full triangular grid. The
selector takes this route when the matrix, or a region of it, is dense enough
that tracking or skipping individual columns within a tile would save no
further work.

\paragraph{Semisparse route.}
Each tile stores only its \emph{active} columns, the columns structurally
nonzero within that block (Section~\ref{sec:format}). The same four cases run,
but each kernel works on the compact active sub-block and scatters its result
into the active layout of the destination tile, and a diagonal tile is held as
a banded factor of half-width $\texttt{kd}$ and factored by a banded Cholesky.
The selector takes this route for structured-sparse matrices in which most
columns of a tile are empty, so that each kernel does work proportional to the
active set rather than to the full $n_b^2$ block.

The two are not separate solvers but two ends of one machinery, and away from
the diagonal they coincide exactly: a fully active tile on this route is a dense
$n_b \times n_b$ block whose active-column map is the identity, so its ~\textsc{syrk} and
~\textsc{gemm} are the dense calls with the scatter a no-op. The diagonal is the one
place where the routes keep their own storage and their own kernel. The dense
route holds it as an upper triangle and factors it with an unbanded~\textsc{dpotrf}, while the semisparse route holds it banded-packed and factors
it with~\textsc{dpbtrf}, which at full bandwidth ($\texttt{kd} = n_b - 1$)
reaches the same factor through a different routine and carries an unused
corner; the panel solve inherits the distinction, running as~\textsc{dtrsm}
against a triangular diagonal tile and as~\textsc{dtbtrs} against a banded one.
Shared without qualification is everything above the kernels: the tiling, the
task list, its per-thread partition, and the dynamic-scheduler-free executor.
The limit is one of the whole matrix, not of a single tile: a matrix in which
\emph{every} tile is fully active never runs on the semisparse route, since
the selector sends anything that dense to the dense tiling outright. Fully
active \emph{tiles}, by contrast, are common on the semisparse route, and are
exactly the no-op-scatter case that a dense border produces
(Section~\ref{sec:grid-alignment}). The four kernels (Cases 1--4) act on the active
sub-blocks through the scatter of Section~\ref{sec:scatter}, driven by the
executor of Section~\ref{sec:executor}.

So that the executor does no run-time scheduling, a single pre-numerical pass,
run once per sparsity pattern, builds the per-tile metadata, the flat task
list, and the per-task scatter that drive
Algorithm~\ref{alg:left-looking-semisparse}.

\paragraph{Per-tile metadata.}
With the grid overlaid on $L$, two parallel passes compute the metadata record
of Section~\ref{sec:format}. One sweep records, for each tile, its active
columns and two fast-path predicates: whether the active columns form a
contiguous range, and whether the tile is full width. For each diagonal tile
the per-tile CSC fill descriptor is scanned to compute
\[
  \texttt{kd}_t \;=\; \max_{(i,j)\,\in\, \text{fill}(t),\, i \le j}\, (j - i),
\]
the LAPACK upper-banded half-width that fixes its storage shape at
$(\texttt{kd}_t + 1) \times n_b$. Tiles are independent, so both passes run in
parallel.

\paragraph{Task list.}
The $(k, m, n)$ iteration space of
Algorithm~\ref{alg:left-looking-semisparse} is enumerated once by a parallel
collector, which skips any task touching a structurally zero tile
($|\mathcal{A}| = 0$) so that the list holds exactly the numerically nonzero
work. Each task carries its routine, its $(k,m,n)$ coordinates, and the three
tiles' resolved buffer offsets. The tasks are then partitioned into per-thread
slices in panel order, so each thread walks a static sequence the executor's
progress matrix can rely on.

\paragraph{Per-task scatter.}
A symbolic pre-pass over the task list records, for each task, the
active-column scatter map it will write through (Section~\ref{sec:scatter}),
so that at numerical time each kernel is a BLAS-3 call followed by a scatter,
with no branching on tile metadata. The whole pre-numerical pass costs
$O(\texttt{nnz}(L))$ and takes a fraction of one numerical factorization, so
it is amortized over the repeated factorizations the design targets;
Section~\ref{sec:experiments} reports the measured breakeven point.

\subsection{Executor and progress matrix}\label{sec:executor}

The symbolic phase produces the task list once, and each thread executes a
static slice of it in order. The partition is by column panel: every task that
writes to one column (its~\textsc{syrk} and~\textsc{gemm} updates, its ~\textsc{potrf}, and its~\textsc{trsm}
solves) lands in a single thread's slice, ordered so that the updates into a
tile precede the task that consumes it. Dependencies then fall into two kinds.
An update and its consumer in the same panel are ordered by the slice itself,
so~\textsc{potrf} and~\textsc{trsm} never wait on their own panel's updates. A read across
panels, where a kernel consumes a tile another column produced, is resolved
through a two-dimensional progress matrix $P$ indexed by tile coordinates: the
producing~\textsc{potrf} or~\textsc{trsm} sets a completion flag on the tile it finalizes, and
the cross-panel consumer spin-waits on that flag before it runs. Because the
task list, its panel partition, and the per-task scatter maps are all fixed at
symbolic time, the executor does no dynamic scheduling at run time. This has
two consequences. First, a single heavy panel, a wide dense border for
example, serializes its own updates on one thread and caps parallelism there,
an Amdahl limit that no scheduling can remove. Second, the supernodal mode, whose updates into a cell arrive from several panels rather than one, cannot rely on in-slice ordering for those updates. Its executor (Section~\ref{sec:supernodal}) is built to this same design and keeps a progress matrix of the same kind for the single-producer dependencies (a cell finalized by one~\textsc{potrf} or one~\textsc{trsm}, where a completion flag is enough) and adds, alongside it, a per-cell counter of outstanding updates that each update decrements and the consumer waits to reach zero. The counter expresses what a flag cannot, namely that all of a cell's writers are done, and it gives the same guarantee without requiring single-thread ownership of a column.
Algorithm~\ref{alg:exec} gives the per-thread loop for the tile modes.

\begin{algorithm}[t]
\caption{Semisparse Cholesky executor, per-thread loop.}
\label{alg:exec}
\small
\begin{algorithmic}[1]
\State \textbf{Input:} task slice $[\text{start}, \text{end})$;
       per-tile metadata $\{\texttt{sa}, \texttt{aind}, \texttt{upper\_bw}\}$;
       scatter array \texttt{scatter\_index}; progress matrix $P$
\For{$\mathrm{idx} \gets \text{start}$ \textbf{to} $\text{end}-1$}
    \State $(\mathrm{op}, m, k, n, i_1, i_2, i_3) \gets \texttt{tasks}[\mathrm{idx}]$
           \Comment{$i_1,i_2,i_3$ index the operand tiles in the tile store}
    \If{$\mathrm{op} = $ \textsc{potrf}}
           \Comment{\textsc{syrk} updates into $L_{kk}$ are earlier tasks in this slice}
        \State $L_{kk} \gets \textsc{DPBTRF}(L_{kk},\, \texttt{kd})$
               \Comment{banded factor in place}
        \State $P[k,k] \gets 1$
               \Comment{finalize $(k,k)$; cross-panel readers wait on this}
    \ElsIf{$\mathrm{op} = $ \textsc{syrk}}
        \State \textbf{wait until} $P[k,n] = 1$ \Comment{cross-panel read of $L_{kn}$}
        \State $L_{kk} \mathrel{-}= L_{kn} L_{kn}^{\top}$
               \Comment{BLAS-3}
    \ElsIf{$\mathrm{op} = $ \textsc{trsm}}
           \Comment{\textsc{gemm} updates into $L_{mk}$ are earlier tasks in this slice}
        \State \textbf{wait until} $P[k,k] = 1$ \Comment{cross-panel: $(k,k)$ finalized}
        \State $L_{mk} \gets \texttt{dtbtrs}(L_{kk},\, \texttt{kd},\, L_{mk},\,
               |\mathcal{A}_{mk}|)$
               \Comment{banded solve on the stored $L_{kk}^{\top}$}
        \State $P[m,k] \gets 1$
    \ElsIf{$\mathrm{op} = $ \textsc{gemm}}
        \State \textbf{wait until} $P[k,n] = 1$ \textbf{and}
               $P[m,n] = 1$ \Comment{cross-panel reads of $L_{kn}, L_{mn}$}
        \State $\mathrm{off} \gets \texttt{scatter\_index}[\mathrm{idx}]$
        \State $L_{mk} \mathrel{-}= L_{mn} L_{kn}^{\top}$
               \Comment{BLAS-3 then scatter}
    \EndIf
\EndFor
\end{algorithmic}
\end{algorithm}

\subsection{The non-uniform supernodal tiling}\label{sec:supernodal}

For the matrices the selector routes to the ultra-sparse end
(Section~\ref{sec:selector}), no fixed tile grid pays off: active-column
masking still leaves each tile almost empty, and the per-tile metadata
outweighs the arithmetic it guards. Here the factorization switches to a
supernodal-sparse layout that replaces the uniform grid with a
\emph{non-uniform tiling} fitted to the symbolic fill, its tiles aligned to
supernode boundaries rather than to a fixed grid. The numerical method it runs
is the established supernodal left-looking Cholesky of sparse direct solvers,
built from an elimination tree, relaxed supernodes, and dense per-supernode
kernels, the same well-understood family that modern solvers such as
symPACK~\cite{jacquelin2016sympack} build on. We take the method as given;
what is ours is its \emph{scheduling}. We run it on a sibling executor built
to the same design as the tile modes' (Section~\ref{sec:executor}), a
shared-memory schedule of static per-thread slices fixed at symbolic time, as a
first-class mode rather than a separate solver. A distributed solver like
symPACK realizes the method through a UPC++/MPI runtime; we carry it with no
distributed-runtime dependency, scheduled once ahead of time and replayed. On
the single-node target this route leads the sparse regime over the
specialized solvers in the repeated-factorization setting
(Section~\ref{sec:experiments}). It shares the selector, the ordering pipeline, and the BLAS backend with the tile modes: a third representation within one framework, its sibling task-DAG executor specialized for non-uniform supernodal tiles instead of
uniform grid tiles.

Integrating the method this way, rather than dispatching the ultra-sparse
matrices to an external solver, pays off twice. First, the once-per-pattern
ordering and symbolic analysis are shared across the thousands of repeated
factorizations and reused unchanged by the log-determinant, with no external
solver to re-analyze the pattern. Second, it avoids external and
distributed runtime overhead on the single-node target: no separate process,
no data marshaling across an interface, no message-passing layer. We compare
against MUMPS, PaStiX, CHOLMOD, and PARDISO~\cite{onemkl_pardiso} (the closest single-node baselines being shared-memory CHOLMOD and PARDISO), and report symPACK as a representative supernodal solver with the caveat, noted above, that it targets
distributed memory (Section~\ref{sec:experiments}).

\paragraph{Non-uniform tiles.}
In place of the uniform $n_b \times n_b$ grid tiles, the supernodal tiling
stores $L$ as variable-sized dense tiles, the \emph{cells} of the supernodal
layout, one per supernode pair $(I, J)$ with $J \ge I$: column supernode $I$
supplies the tile's width, and row supernode $J$ supplies the contiguous run
of $I$'s row pattern that it owns. Each tile is fully dense, and sparsity is
captured by \emph{which} tiles exist rather than by masking within them. All
tiles share one contiguous arena, grouped by column supernode, so a panel of
$L$ is cache-contiguous. These tiles range from fat relaxed supernodes down to
a single column at the irregular end, where the supernodal mode degrades
gracefully toward column-level work instead of switching to a separate scalar
kernel.

\paragraph{Symbolic phase.}
The supernodal tiling reuses the fill-reducing ordering already chosen for the
matrix (Section~\ref{sec:ordering}). From that ordering it builds the
elimination tree of the permuted matrix~\cite{liu1990etree}, post-orders it,
computes column counts~\cite{gilbert1994counts}, forms fundamental
supernodes~\cite{liu1993supernodes}, and relaxes them by the standard
amalgamation heuristic~\cite{ashcraft1989relaxed}, emitting a packed
per-supernode row pattern. This is the conventional supernodal symbolic
pipeline with standard relaxation parameters.\footnote{Ashcraft--Grimes amalgamation thresholds, the same
defaults as symPACK, left untuned for fairness.} What is specific to us is that it reuses the solver's own
fill-reducing ordering, so the analysis is computed once and amortized over
the repeated factorizations, exactly as for the fixed-grid routes.

\paragraph{Numerical phase.}
The numeric work is the left-looking supernodal sweep, expressed in the same
three task kinds the executor already runs: a \textsc{factor} (\texttt{POTRF})
on each diagonal tile $(I,I)$; a \textsc{trsm} of every off-diagonal tile
$(J,I)$ against it; and an \textsc{update} $L_{KJ} \leftarrow L_{KJ} - L_{KI}
L_{JI}^{\top}$ that reconciles the row sets of the two source tiles through a
sorted-subset scatter into the destination, the non-uniform-tile analogue of
the grid-tile scatter of Section~\ref{sec:scatter}. These tasks are scheduled
just as in the tile executor (Section~\ref{sec:executor}): a static task DAG,
balanced by the same load-balancing scheme and synchronized by the same two primitives, the progress matrix for the single-producer dependencies and the per-cell update counters for the cross-panel ones. The only difference is that the scheduled unit is a
non-uniform supernodal tile rather than a uniform grid tile.

\section{Experiments}\label{sec:experiments}

We evaluate the structure-adaptive tile factorization, meaning the per-matrix
selector together with its three representations, on the benchmark suite of
Section~\ref{sec:patterns}, comparing against
MUMPS~\cite{amestoy2001fully}, PaStiX~\cite{henon2002pastix},
PARDISO~\cite{onemkl_pardiso}, CHOLMOD~\cite{chen2008algorithm}, and
symPACK~\cite{jacquelin2016sympack}. The question is not whether any one
representation is best, but whether routing each matrix to the right one beats
committing to a fixed structure. We find that the selector outperforms every
uniform mode in aggregate, with the active-column tile carrying the
structured-sparse regime, a dense tiling the near-dense matrices, and the
supernodal tiling the irregular ones, and no competitor leading across all
three regimes. We report storage, single-factorization time, the effect of the fill-reducing ordering, the
quality of the routing decision, strong scaling, and preprocessing
amortization (Section~\ref{sec:amortization}).

\subsection{Experimental setup}\label{sec:setup}

All experiments run on the KAUST Ibex cluster. The main comparison uses an
Intel node (two $20$-core Intel Xeon Gold CPUs, $40$ cores, $384$~GB),
and a second AMD node (two $64$-core AMD EPYC CPUs, $128$ cores, $512$~GB) checks portability to a second architecture. Each sweep reserves
every core on its node, so on neither machine can a co-tenant job compete for
memory bandwidth. On each
node every solver is built with the same Intel oneAPI compiler and
optimization level and links the same Intel MKL~$2025.3$ for BLAS-3 and banded
LAPACK, so any performance difference comes from the factorization algorithm
and not the build.

For sTiles, the per-matrix selector (Section~\ref{sec:selector}) fixes the
tile mode from the symbolic fill before any numerical work. The tile size is
not tuned per matrix. Following the cache-aware sizing of the original
framework~\cite{abdulfattah2025stiles}, $n_b$ is set so that a tile and its
operands fit the level-3 cache: $n_b = 80$ by default, and $n_b = 120$ in the
structured-sparse regime, where the larger tile better amortizes the
active-column metadata. The competitors (MUMPS~5.7.3, PaStiX~6.4.0, CHOLMOD from
SuiteSparse~7.8.3, PARDISO as shipped in Intel
oneMKL~\cite{onemkl_pardiso}, and symPACK) are configured as in
Table~\ref{tab:solvers}, each with its own fill-reducing ordering and
single-node parallelism at its recommended defaults. We did not sweep
competitor block sizes or scheduler parameters, so their times are a
default-configuration baseline rather than a per-matrix-tuned ceiling, and
sTiles' own $n_b$ is likewise the fixed cache rule above rather than a
per-matrix search.

Held fixed across solvers are the matrix, double precision, and exact factorization. With the ordering held fixed, the three routes produce the same factor
entry by entry: across the matrices we checked, the largest entrywise
difference between the dense, active-column and supernodal factors is
$\langle$value$\rangle$, so the route decides speed and memory and nothing
else. Each reported time is the minimum of three numerical
factorizations after a warm-up, taken over the swept core counts
($\{1,2,4,8,16,32,40\}$ on Intel and $\{1,2,4,8,16,32,64,128\}$ on AMD). We
report the numerical factorization only, since the analysis is computed once
per pattern and amortized (Section~\ref{sec:amortization}). The comparison is
therefore end-to-end within that phase: a reported time includes the fill of
each solver's own ordering, not the kernels alone.

\begin{table}[t]
\centering\small
\caption{Competitor solver configurations; the shared environment is given in
  Section~\ref{sec:setup}. In the ordering column, \texttt{auto} marks the solvers that select a
  fill-reducing ordering automatically (sTiles, MUMPS, CHOLMOD) and ND a fixed
  nested dissection (PARDISO, PaStiX, symPACK); the resulting fills are
  comparable across solvers.}
\label{tab:solvers}
\begin{tabular}{@{}llll@{}}
\toprule
solver & version & factorization & ordering \\
\midrule
sTiles  & this work         & adaptive tile (dense / semi / supernodal) & auto \\
PARDISO & oneMKL~2025.3     & supernodal       & ND \\
MUMPS   & 5.7.3             & multifrontal     & auto \\
PaStiX  & 6.4.0             & supernodal       & ND \\
CHOLMOD & SuiteSparse~7.8.3 & supernodal       & auto \\
symPACK & GitHub build      & supernodal       & ND \\
\bottomrule
\end{tabular}
\end{table}

\subsection{Overall comparison}\label{sec:overview}

\begin{figure*}[t]
  \centering
  \includegraphics[width=\linewidth]{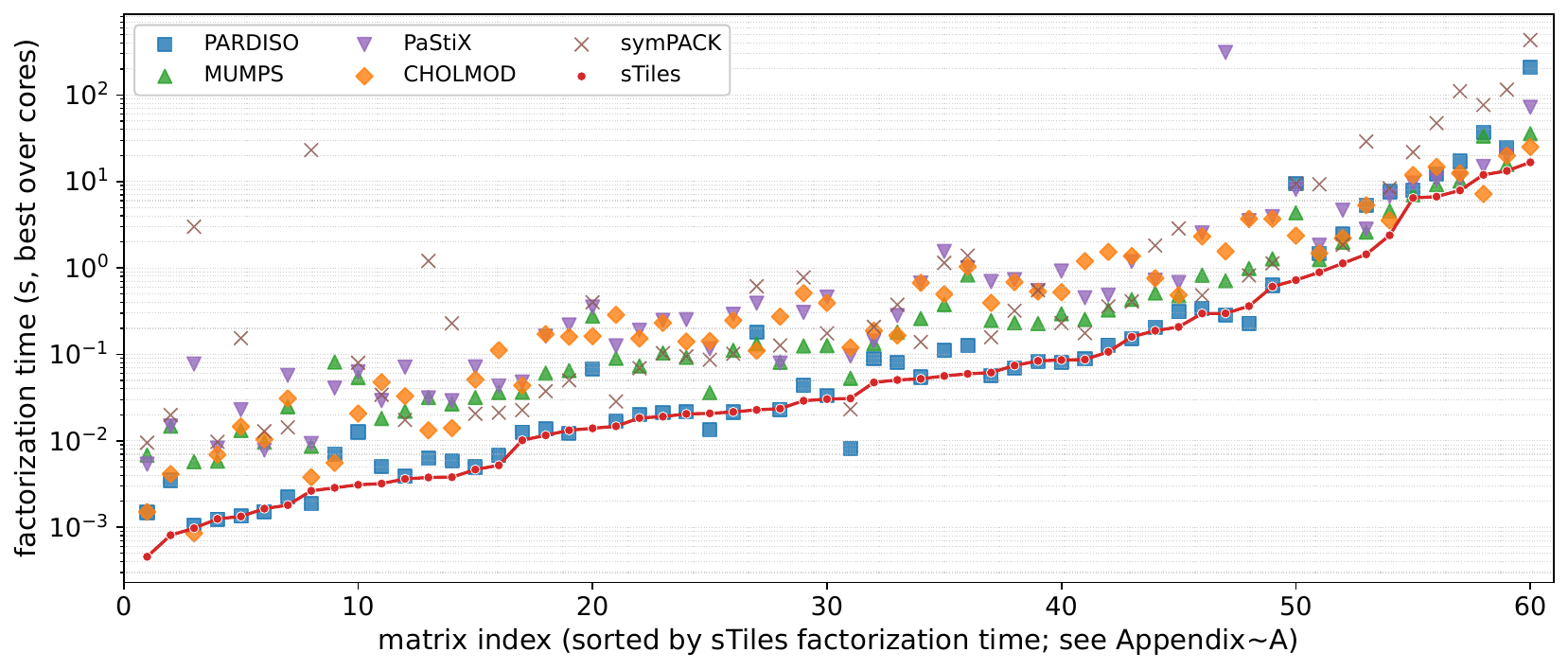}
  \caption{Best factorization time (minimum over the swept core counts) for
    every matrix in the $60$-matrix suite and every solver, on the Intel
    node, and numbered as in Appendix~\ref{app:matrices}. Matrices are
    sorted along the horizontal axis by sTiles' time, so
    the sTiles curve (red) is monotonic; a competitor marker below it is
    faster than sTiles on that matrix, above it slower. PARDISO is the
    closest solver: it is faster than sTiles on $15$ matrices, all below
    $0.5$~s; sTiles is faster on the other $45$, including all $12$
    above $0.5$~s (\#49--60).
    MUMPS, PaStiX, CHOLMOD, and symPACK are slower than sTiles on $97$ to
    $100\%$ of the matrices each of them factors. The vertical axis is
    logarithmic.}
  \label{fig:all-matrices}
\end{figure*}

The per-matrix figures and the regime breakdown in this section are on the Intel node (Section~\ref{sec:setup}); Table~\ref{tab:byclass} also reports the AMD EPYC results, which we summarize at the end of the section. Figure~\ref{fig:all-matrices} plots every solver's best factorization time on every matrix, ordered by sTiles' time. We read it per matrix before summing, because a user runs one model, and hence one sparsity pattern, not the whole suite. On a per-matrix basis PARDISO is the only broadly competitive solver: against MUMPS, CHOLMOD, PaStiX, and symPACK, sTiles is faster on $97$ to
$100\%$ of the matrices each of them factors on the Intel node, by a geometric mean of $4.1$ to $8.9\times$. Against PARDISO the margins are much narrower, though sTiles remains ahead on every summary statistic. Writing the per-matrix ratio as PARDISO's factorization time over sTiles' (so that a value above $1$ means sTiles is faster), sTiles factors $45$ of the $60$ faster, with a median of $1.19\times$ and a geometric mean of $1.50\times$. 

\begin{figure}[t]
  \centering
  \includegraphics[width=0.78\linewidth]{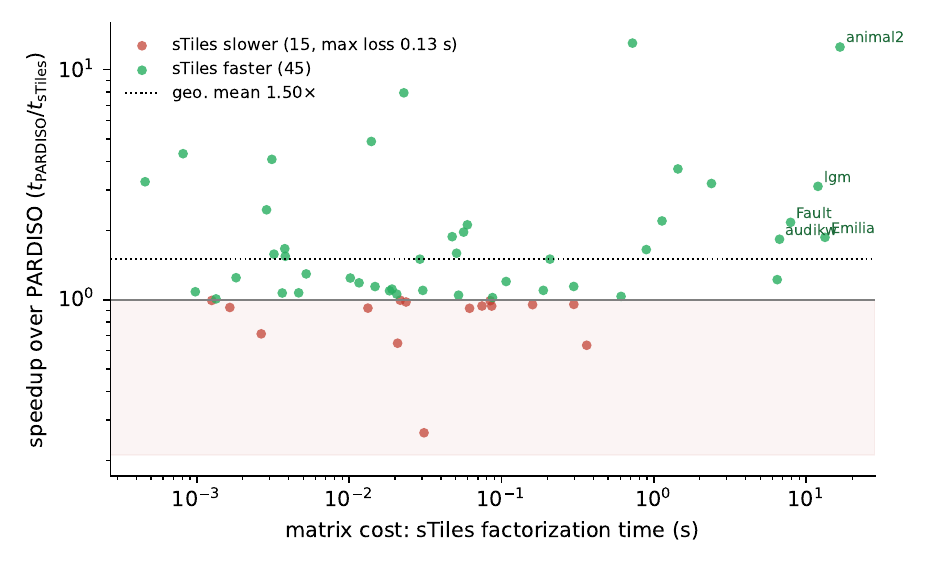}
  \caption{Per-matrix speedup of sTiles over PARDISO against matrix cost
    (both axes logarithmic; a value above the $1\times$ line means sTiles is
    faster). sTiles is faster on $45$ matrices (green) and slower on $15$
    (red), with a geometric-mean ratio of $1.50\times$. The two outcomes are
    asymmetric: every loss is on an inexpensive matrix, all $15$ under
    $0.15$~s and the largest margin $0.133$~s, while the wins are on the
    expensive matrices and reach minutes.
    The aggregate advantage comes from the upper right, the few costly
    factorizations where the representation choice matters.}
  \label{fig:speedup-dist}
\end{figure}

What makes the aggregate margin so much larger than that median is an
asymmetry between the two outcomes (Figure~\ref{fig:speedup-dist}). Where
sTiles is slower than PARDISO, it is slower on an inexpensive matrix and by a
margin measured in milliseconds: every one of the $15$ such matrices is under
$0.15$~s, the median difference is $5$~ms, and all $15$ together sum to
$0.20$~s. The largest absolute margin, \texttt{net1628760}, is $0.133$~s
($0.36$ against $0.23$~s), and the largest relative difference,
\texttt{83o4NNNo} at $0.26\times$, is $31$ against $8$~ms. Where sTiles is
faster, it is faster on an expensive matrix and by a large margin, up to
$192$~s on \texttt{animal2}, with the $45$ summing to $265$~s. Which case a
given matrix falls into is not known before it is factored. The $4.7\times$
aggregate (Table~\ref{tab:byclass}) is a consequence of this asymmetry rather
than a separate claim, since a sum across matrices of vastly different cost is
set by the few expensive ones, which is where the representation choice has
the most effect. The times span six orders of magnitude, making the
structural range of Section~\ref{sec:patterns} concrete.

Table~\ref{tab:byclass} splits the same comparison into the three groups
the router assigns each matrix to, sparse, semisparse, and dense.

\begin{table}[t]
\centering
\caption{Aggregate factorization time (s) per solver, split by the three
groups the router (selector) assigns each matrix to, on the \textbf{Intel}
node (best over cores $\{1,\dots,40\}$) and the \textbf{AMD}~EPYC node (cores
$\{1,\dots,128\}$); the routing is identical on both. The \emph{matrices} row
is how many the router sends to each group. Values in parentheses are how many
times slower than sTiles, per group and overall in the total column; the
\emph{wins} column counts how many of the $60$ matrices each solver factors
faster than sTiles.
\textsuperscript{*}symPACK does not factor two matrices on Intel
(\texttt{pedigree}, \texttt{yU0G1u}) and three on AMD (those two plus
\texttt{animal2}), so its columns sum the matrices it completes and its
ratios, and its wins, are taken against sTiles on the same subset.}
\label{tab:byclass}
\small
\begin{tabular}{lrrrrr}
\toprule
 & \multicolumn{4}{c}{\textbf{Router}~~{\footnotesize(factorization time in seconds (speedup))}} & \\
\cmidrule(lr){2-5}
solver & sparse & semisparse & dense & total & wins \\
matrices & $47$ & $7$ & $6$ & $60$ & \\
\midrule
\multicolumn{6}{@{}l}{\textit{Intel Xeon Gold ($40$ cores)}} \\
sTiles  & $\mathbf{51.6}$ & $\mathbf{0.11}$ & $\mathbf{21.0}$ & $\mathbf{72.6}$ & -- \\
PARDISO & $109.0\ (2.1\times)$ & $0.26\ (2.4\times)$  & $228.6\ (10.9\times)$ & $337.9\ (4.7\times)$ & $15$ \\
MUMPS   & $88.0\ (1.7\times)$  & $1.39\ (12.6\times)$ & $47.1\ (2.2\times)$ & $136.4\ (1.9\times)$ & $0$ \\
CHOLMOD & $95.3\ (1.8\times)$ & $1.73\ (15.8\times)$ & $33.8\ (1.6\times)$ & $130.8\ (1.8\times)$ & $2$ \\
PaStiX  & $401.2\ (7.8\times)$ & $1.78\ (16.2\times)$ & $93.5\ (4.5\times)$ & $496.5\ (6.8\times)$ & $0$ \\
symPACK\textsuperscript{*} & $448.3\ (8.7\times)$ & $2.67\ (25.0\times)$ & $453.6\ (21.6\times)$ & $904.6\ (12.5\times)$ & $1$ \\
\midrule
\multicolumn{6}{@{}l}{\textit{AMD EPYC ($128$ cores)}} \\
sTiles  & $\mathbf{64.5}$ & $\mathbf{0.19}$ & $\mathbf{19.1}$ & $\mathbf{83.7}$ & -- \\
PARDISO & $85.9\ (1.3\times)$ & $0.34\ (1.8\times)$ & $166.3\ (8.7\times)$ & $252.5\ (3.0\times)$ & $5$ \\
MUMPS   & $147.3\ (2.3\times)$ & $1.63\ (8.5\times)$ & $72.6\ (3.8\times)$ & $221.6\ (2.6\times)$ & $0$ \\
CHOLMOD & $171.6\ (2.7\times)$ & $1.56\ (8.1\times)$ & $46.7\ (2.5\times)$ & $219.9\ (2.6\times)$ & $2$ \\
PaStiX  & $203.1\ (3.1\times)$ & $1.95\ (10.1\times)$ & $77.4\ (4.1\times)$ & $282.5\ (3.4\times)$ & $1$ \\
symPACK\textsuperscript{*} & $680.8\ (10.6\times)$ & $1.97\ (10.5\times)$ & $32.7\ (5.0\times)$ & $715.4\ (10.1\times)$ & $4$ \\
\bottomrule
\end{tabular}
\end{table}

Reading down its columns, Table~\ref{tab:byclass} shows that each of the three
groups is fastest under its matching tiling and that no competitor is strongest in more than one regime. The aggregate above does not reveal where the
advantage comes from; decomposing it by regime does.

\paragraph{Sparse (47 matrices): the established regime.}
The irregular and general-sparse matrices (finite-element fronts, networks,
graph precisions) route to the supernodal tiling. This is the regime mature
sparse solvers are built for: PARDISO factors $15$ of the $47$ faster than
sTiles, which are all of its suite-wide wins, on the many inexpensive graphs.
sTiles nonetheless leads the regime total ($52$~s, against MUMPS $88$~s and
PARDISO $109$~s), because it is faster on the expensive factorizations that
dominate the sum, Emilia\_923 by $1.9\times$, Fault\_639 by $2.2\times$,
audikw\_1 by $1.8\times$, and bone010 by $1.2\times$. The supernodal route is
therefore comparable to the established solvers on the matrices they target,
and ahead on the heavy tail, which is what we would expect.

The $15$ slower cases divide into two causes, neither of them the kernel. Among
the five INLA graphs, sTiles is faster single-threaded (median $1.14\times$)
and forms less fill ($0.83\times$), and is slower only because it scales worse
($0.73\times$ of PARDISO's speedup); \texttt{83o4NNNo} is the extreme, ahead
$2.6\times$ on one core but reaching $1.7\times$ on forty against PARDISO's
$17\times$, because the static partition cannot expose enough independent
work on an ultra-sparse irregular graph. On the ten finite-element matrices
the pattern inverts: sTiles scales better ($1.14\times$) at essentially
identical fill, but is $18\%$ slower per core, and the differences are
correspondingly narrow (median $0.95\times$). Since both solvers dispatch the
same BLAS-3 through the same MKL and form comparable factors, neither gap is
one a faster kernel would close.

\paragraph{Semisparse (7): the regime with no standard representation.}
This is the structured middle: bordered and banded latent-Gaussian
precisions, dense enough that level-3 BLAS pays yet far too sparse to store
dense. sTiles is fastest on all seven, by $2.4\times$ over the closest competitor (PARDISO) and $13$ to $25\times$ over the rest. These are wide relative margins on inexpensive factorizations. The whole regime totals under a tenth of a second (Table~\ref{tab:byclass}),
because structured banded and bordered matrices are inexpensive to factor, so
it weighs little in the \emph{single-factorization} aggregate even though its
margins are the largest. That aggregate is the wrong lens for this regime. A
single INLA fit reuses one precision pattern the thousands of times the
workload issues, and at that scale the margin compounds: the regime's
$0.15$~s per-factorization edge over PARDISO projects to roughly $150$ to
$1{,}500$~s of wall-clock across the thousands of reuses in a fit. This is a
factorization-time projection from the per-factorization edge and the reuse
count, not an end-to-end INLA timing. (A full fit also forms posterior
marginals through selected inversion, which this solver does not provide; that
component is developed separately~\cite{abdulfattah2025selinv}.) A single
model (\texttt{bern}) accounts for tens to $\sim\!150$~s of that per fit.
Under the reuse the design targets, it is this regime, not the expensive tail,
where the per-fit time goes, and the suite total hides the semisparse
contribution precisely because it ignores that reuse. This is the gap the
active-column tile is built for: it strips each tile to its active columns while
keeping every kernel a dense BLAS-3 call, capturing the structure without
paying for the zeros, and its margin widens with size. The result is scoped to
the structures that align with the fixed grid, a regular band of bounded width
plus a dense border, which fills a predictable set of active columns per tile.
These seven matrices are exactly those: the banded \texttt{sem\_n} series, the
bordered \texttt{bern}, the arrowhead \texttt{pedigree}, and the
spatio-temporal \texttt{stcov}, all latent-Gaussian.

\paragraph{Dense (6): routing by density, not fill.}
The near-dense matrices (animal models, three \texttt{lgm} precisions, and
\texttt{lidense}) route to the dense tiling, where sTiles is fastest, ahead of PARDISO by $10.9\times$ in
the regime total and by $12.6\times$ on the $50{,}126^2$ \texttt{animal2}
($16.6$~s against $208.8$~s). What stands out here is a change of order among
the competitors. PARDISO, the closest elsewhere, is at $229$~s here, behind
CHOLMOD ($34$~s), MUMPS ($47$~s), and PaStiX ($94$~s), because its sparse path carries the explicit zeros of a near-full front rather
than switching to dense kernels. This is the empirical reason the selector
routes a near-dense matrix by its density rather than by its fill ratio alone,
which would send it to a sparse method. The point is structural, that a
near-dense front belongs on dense kernels, and we make it on six matrices from
three structures (the animal models, three \texttt{lgm} precisions, and the
dominant-dense-block \texttt{lidense} precision) rather than as a ranking of
dense solvers at large. The \texttt{\_bw} suffix in the \texttt{lgm} names does
not track block occupancy: three such precisions route here and three route to
the sparse regime.

Across the three regimes, the leading competitor changes each time: MUMPS on
sparse, PARDISO on semisparse, CHOLMOD on dense, while sTiles is first in all
three. The structural spectrum of Section~\ref{sec:patterns} is thus
operational and not merely descriptive: the right representation changes with
the matrix, and any solver committed to one structure is left behind somewhere
on the suite. The sparse regime, broad at $47$ matrices, is where the
competitors are strongest.

These speedups cost essentially nothing in stored factor size. The resident
factor sTiles stores ($\mathrm{nnz}(L)$ at fp64, summed over the suite,
$69$~GB) is within a few percent of the supernodal and multifrontal
competitors ($0.99\times$ CHOLMOD, $1.05\times$ PARDISO, $1.07\times$ MUMPS, $1.15\times$ PaStiX). This spread is set by each solver's fill-reducing
ordering rather than by the storage format: the active-column tile reaches
level-3 BLAS efficiency while holding essentially the same $\mathrm{nnz}(L)$
as a conventional sparse factor.

\paragraph{AMD EPYC.}
The lower panel of Table~\ref{tab:byclass} repeats the comparison on the AMD
node, under the same routing. The ranking is unchanged: sTiles is first in
every regime and leads the total by $2.6$ to $3.4\times$ over the direct
solvers and by $10.1\times$ over symPACK, with PARDISO's dense-regime
disadvantage ($8.7\times$) intact. The margins are narrower than on Intel,
because the EPYC's memory bandwidth throttles every solver on the largest
factorizations and compresses the spread rather than reordering it. The
narrowing is a hardware effect and not a measurement artifact: the AMD sweep,
like the Intel sweep, ran on a dedicated whole-node allocation, so the
compressed spread reflects the EPYC's bandwidth ceiling rather than
contention. The per-matrix split also shifts with the architecture: PARDISO
beats sTiles on $5$ matrices on AMD against $15$ on Intel, and symPACK on $4$
against $1$, while the aggregate ordering is unchanged. The ordering of the
solvers, and sTiles' lead in every regime, carries to the second architecture.

\subsection{Controlling for the fill-reducing ordering}

Automatic ordering selection is a property of the solver, not a separate
advantage we grant sTiles, and a controlled test confirms that it does not
explain the heavy-tail lead. Pinning both sTiles and PARDISO to METIS on the
large factorizations that carry the sparse lead, and re-timing, sTiles factors
\texttt{Fault\_639}, \texttt{audikw\_1}, and \texttt{bone010} faster by $1.65$
to $1.83\times$. The two solvers' METIS orderings differ in fill by under
$10\%$, so even at matched fill the lead is $1.5$ to $1.6\times$. These are
the METIS-matched margins; the per-matrix ratios in
Section~\ref{sec:overview} and the appendix differ because there each solver
runs its own best ordering. That the lead holds both ways, under a common
ordering and under each solver's own, is the robustness we are after. The
heavy-tail margins therefore reflect the numerical factorization phase rather
than a lower-fill ordering. This control isolates the ordering; it does not by
itself separate the schedule from the kernels or from the shared-memory
execution within that phase.

\subsection{Does the router choose well?}
\label{sec:routing-accuracy}

The aggregate gains of Section~\ref{sec:overview} rest on the router sending
each matrix to the right mode. Three tests bear this out: selection beats any
fixed mode, it picks the fastest mode wherever time is at stake, and it is
insensitive to the exact thresholds.

\paragraph{Selection beats every fixed mode.}
Forcing one mode across the whole suite (Table~\ref{tab:selector}), the
selector ($73.1$~s) beats the best fixed mode, all-sparse, by $1.58\times$,
all-semisparse by $2.07\times$, and all-dense by $2.62\times$. Each fixed mode
pays on the regimes it does not fit: the dense tiling runs everywhere, since
it prunes empty tiles, but carries explicit zeros on the sparse matrices, just
as the sparse and semisparse modes do on the dense ones. No single structure
fits the whole suite, so selection is a first-order design decision rather than
a tie-break. These margins do conflate two adaptations, since the forced modes
run at $n_b = 80$ while the selector also raises $n_b$ to $120$ on structured
matrices. Mode choice is the larger effect, but separating the two cleanly
would require re-running each fixed mode at its own best $n_b$.

\begin{table}[t]
\centering

\caption{Per-matrix selection versus any single fixed mode: total
factorization time forcing one mode across all $60$ matrices (Intel node,
summed best over cores). No fixed mode matches the selector, and each pays
on the regimes it does not fit. Forced sparse and semisparse use
$n_b = 80$; the selector additionally adapts $n_b$, so part of its margin
is the tile-size choice.}
\label{tab:selector}
\begin{tabular}{lrr}
\toprule
configuration & total time (s) & vs selector \\
\midrule
per-matrix selector (auto) & $73.1$  & -- \\
all sparse                 & $115.6$ & $1.58\times$ \\
all semisparse             & $151.5$ & $2.07\times$ \\
all dense                  & $191.4$ & $2.62\times$ \\
\bottomrule
\end{tabular}
\end{table}

\paragraph{Routing accuracy.}
On the $29$ matrices that take at least $0.05$~s in the fastest fixed mode,
the selector picks the empirically fastest of the three on $25$, or $86\%$. All four misses cost at most $1.14\times$
(\texttt{lgm\_50400\_bw2} $1.14\times$, \texttt{lgm\_50000\_bw15000}
$1.09\times$, \texttt{nd12k} $1.02\times$, and \texttt{lgm\_100200\_bw1}
$1.01\times$). Across the full suite, the raw rate is $87\%$, and the further
misses are on matrices of a few tens of milliseconds. A misrouted matrix
degrades gracefully to the next-best mode. The deeper guarantee, though, is not this rate but the supernodal fallback
(Section~\ref{sec:selector}). The accuracy figures bound the \emph{cost} of
the two specialized modes' misfires, while the fallback bounds the \emph{risk}
of a structure the features were never tuned for.

\paragraph{Threshold sensitivity.}
The four cutoffs are round numbers set by hand, which invites the objection
that they are fitted to this suite. They are not. Perturbing each across the
ranges of Table~\ref{tab:sensitivity}, over the $36$ matrices whose features
were logged (including $7$ of the $9$ costliest), moves the aggregate
factorization time by at most $2.3\%$, and the defaults are not even the
per-suite optimum: they sit on a wide plateau. The reason is structural. Near
a boundary, a perturbed cutoff flips a matrix between modes that run at nearly
the same speed on it, so the aggregate barely moves even as the discrete
choice does. The baseline is within $2.0\%$ of the per-matrix oracle ($67.3$ against $65.9$~s). Each cutoff is the round number nearest the point where a representation's cost model (an active-column tile pays only once its blocks are dense enough and
its fill low enough to keep the BLAS-3 kernels efficient), so the suite tests
thresholds it did not set. We deliberately keep four hand-set cutoffs rather
than a decision tree learned over the $60$ labeled matrices: cross-validated
on a sample this small, a classifier would report noisy accuracy and could fit
the suite in the way the coarse rule avoids, and the plateau leaves it nothing
to gain, since near every boundary the competing modes run at nearly the same
speed.
\begin{table}[t]
\centering\small
\caption{Threshold sensitivity. Each routing cutoff (Section~\ref{sec:selector}) is swept individually over the listed range with the others held at their default, and the routing is recomputed offline. The selector's aggregate factorization time over the $54$ matrices whose routing features were logged (including $7$ of the $9$ costliest matrices, those above $1$~s; the remaining matrices route to the supernodal default far from every gate and do not move) changes by at most the last column. The hand-set defaults sit on a wide plateau, not a tuned optimum.}
\label{tab:sensitivity}
\begin{tabular}{@{}lccc@{}}
\toprule
cutoff & default & swept range & max time change \\
\midrule
dense gate $\rho \ge$ & $0.90$ & $[0.80,\,0.99]$ & $2.3\%$ \\
fill gate $\phi \le$ & $2.0$ & $[1.0,\,4.0]$ & $0.5\%$ \\
semi gate $\rho \ge$ & $0.15$ & $[0.05,\,0.50]$ & $0.2\%$ \\
skew gate $\sigma \ge$ & $20$ & $[5,\,100]$ & $0.4\%$ \\
\bottomrule
\end{tabular}
\end{table}

\subsection{Strong scaling}

\begin{figure}[t]
  \centering
  \includegraphics[width=0.72\linewidth]{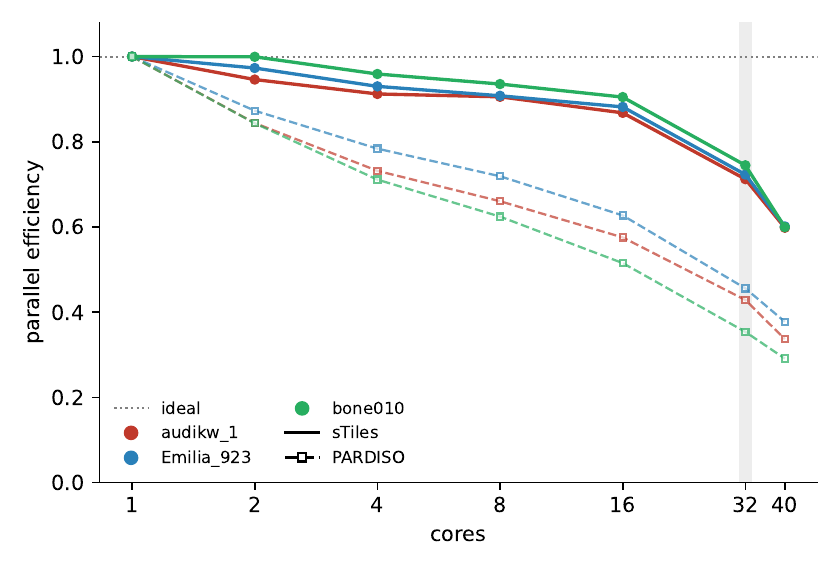}
  \caption{Parallel efficiency (speedup over one core divided by the core
    count; $1.0$ is ideal) of a single factorization on three large
    finite-element matrices, sTiles (solid) against PARDISO (dashed). sTiles
    holds above $0.85$ through $16$ cores, where PARDISO has already fallen to
    $0.5$--$0.6$; both drop sharply past $32$ cores (shaded band), the
    socket's memory-bandwidth ceiling for sparse Cholesky.}
  \label{fig:scaling}
\end{figure}

How a single factorization scales depends on the matrix
(Figure~\ref{fig:scaling}). On the large finite-element matrices that
dominate runtime, sTiles scales well: from $1$ to $40$ cores it speeds up $23$
to $24\times$ (Emilia\_923 $24.1\times$, Fault\_639 $23.3\times$, audikw\_1
$23.9\times$, bone010 $24.0\times$), ahead of PARDISO's $12$ to $15\times$ on
the same matrices. Read as efficiency, sTiles stays above $0.85$ out to $16$
cores while PARDISO falls to $0.5$--$0.6$ over the same range, and the two
bands never cross. Both then drop past $32$ cores; sTiles gains under $8\%$
going from $32$ to $40$ cores, consistent with the socket's memory-bandwidth
ceiling for sparse Cholesky rather than with an algorithmic limit. The small
matrices, by contrast, expose limited intra-factorization parallelism and
reach their best time well below the full node, so we report each matrix at
its best core count. The semisparse route falls entirely in this category. Its
structured precisions all factor in under a tenth of a second, below the size
that rewards the full node, so the figure shows the sparse route, where the
suite's runtime actually concentrates, rather than a semisparse scaling curve
that would flatten at a handful of cores.

\subsection{Preprocessing amortization}\label{sec:amortization}

sTiles buys its inexpensive factorization with the heaviest preprocessing of any
solver we compare (Table~\ref{tab:symbolic}): $787.4$~s of symbolic analysis
across the $60$ matrices at $40$ cores, against $154$ to $206$~s for the
others. The analysis selects the per-matrix mode, runs the ordering
comparison, computes the symbolic fill that fixes each tile's active columns,
and builds the static schedule the executor later replays. What matters,
though, is the total time. Running the analysis once and then $n$ numerical
factorizations of the same pattern costs $T(n) = a + n\,c$. sTiles maximizes
the one-time analysis $a$ but minimizes the per-factorization cost $c$,
whose $72.6$~s aggregate is the least expensive of any solver. This split is exposed in the
interface rather than hidden inside it: the C API and its Python and R
bindings separate a once-per-pattern \texttt{analyze} from a repeated
\texttt{factorize} that reuses it, so a caller pays $a$ once and $c$ per
iteration without managing the symbolic state.

\begin{table}[!ht]
\centering
\caption{Symbolic analysis ($a$) and numerical factorization ($c$) time,
summed across the $60$ matrices (analysis at $40$ cores; factorization the
per-solver aggregate of Table~\ref{tab:byclass}), with the breakeven
factorization count
$n^\star=(a_{\mathrm{sTiles}}-a)/(c-c_{\mathrm{sTiles}})$ beyond which the
sTiles total $a+nc$ is lowest.
\textsuperscript{*}symPACK does not cover the whole suite, so its
$n^\star$ is indicative rather than exact.}
\label{tab:symbolic}
\small
\begin{tabular}{lrrr}
\toprule
Library & Analysis $a$ (s) & Factorization $c$ (s) & Breakeven $n^\star$ \\
\midrule
sTiles  & $787.4$ & $72.6$  & --     \\
PARDISO & $194.6$ & $337.9$ & $2.2$  \\
MUMPS   & $154.4$ & $136.4$ & $9.9$  \\
CHOLMOD & $158.1$ & $130.8$ & $10.8$ \\
PaStiX  & $205.7$ & $496.5$ & $1.4$  \\
symPACK\textsuperscript{*} & $159.7$ & $904.6$ & $0.8$  \\
\bottomrule
\end{tabular}
\end{table}

That breakeven is small (Table~\ref{tab:symbolic}): $n^\star \approx 2$ against
PARDISO, $\approx 1.5$ against PaStiX, $\approx 10$ against MUMPS and
CHOLMOD, and below one against symPACK. The reuse count then decides which of three regimes applies.
A single latent-Gaussian fit issues thousands of factorizations of one
pattern, so $n \gg n^\star$ always, and the analysis amortizes to about $1\%$
of total work at $1000$ reuses and $0.1\%$ at $10{,}000$; here the trade is
decisive. At modest reuse it is thinner and depends on the competitor. A
Newton or interior-point loop of tens of iterations clears the breakeven
against PARDISO and PaStiX but only reaches it against MUMPS and CHOLMOD, and
the analysis still stands at several percent of total work rather than a
negligible fraction, so the advantage is real but slim, not the
amortized-to-nothing case that INLA provides. Below the breakeven, on a
one-off solve or a loop shorter than $n^\star$, sTiles loses by up to the
analysis gap, and we claim no advantage there.

The cost concentrates sharply: five matrices (\texttt{Emilia\_923},
\texttt{animal2}, \texttt{audikw\_1}, \texttt{Fault\_639},
\texttt{bone010}) account for $59\%$ of the analysis total, and all five form
a factor of roughly a billion nonzeros. The size of $L$ explains $90\%$ of the
variation in analysis time across the suite, which is what one expects of a
pass whose work is the fill it enumerates.

It is also structural, not slack. A phase profile at $40$ cores puts $42\%$ of
the analysis in the exact fill-in passes that score candidate orderings and
$32\%$ in the supernodal symbolic factorization on the winning permutation;
counting active tiles adds a further $10\%$, while everything downstream of
the ordering choice, materializing the pattern of $L$, allocating the factor,
and building the schedule, together stays under $2\%$. Of the candidate
scoring, $74\%$ is spent on permutations that are then discarded, at a median
of four candidates per matrix. The measurement is the transitive closure of
symbolic Cholesky and is sequential within each candidate, parallelizable only
along an elimination tree that collapses to a chain on the banded precisions
the workload targets. That same measurement finds the tile-aligned ordering
behind the inexpensive factor, so it is the price of that factor rather than
overhead. Pruning it when the structure already fixes the best candidate is
left to future work.

\subsection{GPU Extension: Challenges and Opportunities}\label{sec:gpu}



Not every sparse matrix is worth sending to a GPU. There is a trade-off between the cost of host-device data movement and the arithmetic intensity of the resulting operation, which may not amortize the introduced overhead. The natural first step is the dense route. Its tiles can grow on a GPU,
from the $80$ to $120$ that suit a CPU core's cache to $512$ and beyond,
and the kernels it issues, \textsc{potrf}, \textsc{trsm}, \textsc{syrk}
and \textsc{gemm} on dense tiles, are the ones the vendor libraries run
closest to peak. 

We leave out the other two routes out; we are
explicit that the case for the semisparse route made in this paper does
not carry over to a GPU. On the CPU the active-column tile pays because it
turns a sparse tile into one dense BLAS-3 call on a packed sub-block. On a 
device where the same call is small, its cost is dominated by the launch rather
than the arithmetic intensity; the scatter that follows it is irregular memory
traffic, and the tiles of one factor range from nearly full to nearly
empty, so an implementation would have to mix dense and sparse tile
kernels within one schedule and decide per tile which one pays. Whether
that ever beats the CPU is an open question, not a consequence of the CPU
result. The sparse route, with its small irregular fronts, is the regime
where the device is least likely to pay at all.

We implement the dense route on CUDA: one host thread walks the static
schedule of Section~\ref{sec:executor} and issues the tile kernels through
cuBLAS and cuSOLVER to a pool of $16$ streams, with CUDA events in place of
the progress matrix, so that a kernel reading a tile waits on the event
recorded by the kernel that produced it and independent tiles overlap on
different streams. The values are uploaded tile row by tile row while the
factorization of the earlier rows proceeds, and the factor stays on the
device. We ran the six matrices that the router sends to the dense route
on a node with one NVIDIA A100 GPU (80~GB), and
compare against sTiles on the two CPU nodes of Section~\ref{sec:setup}. Both sides use the ordering the library
itself selects, and the CPU side runs in the automatic mode of
Section~\ref{sec:selector}, so the baseline is the solver of this paper and
not a configuration chosen for the device. Correctness is checked against a CPU run on the GPU's own host, which agrees
on the log-determinant and the solve on all six.

\begin{table}[t]
\centering
\caption{The six dense-route matrices on one NVIDIA A100 GPU against sTiles
on the Intel and AMD nodes of Section~\ref{sec:setup} (seconds, best of
three). \emph{Movement} is the transfer both ways and \emph{numeric} the
factorization without it, so \emph{total} is the cost when the caller wants
the factor on the host; the last two columns keep the factor on the device,
upload overlapped and no return. Both speedups are over the faster CPU node,
in bold. Log-determinants agree to a relative $10^{-14}$.}
\label{tab:gpu}
\small
\setlength{\tabcolsep}{4pt}
\begin{tabular}{lrrrrrrrrr}
\toprule
 & & \multicolumn{2}{c}{CPU} & \multicolumn{4}{c}{GPU, factor returned} & \multicolumn{2}{c}{GPU, factor kept} \\
\cmidrule(lr){3-4} \cmidrule(lr){5-8} \cmidrule(lr){9-10}
matrix & $n$ & Intel & AMD & movement & numeric & total & speedup & time & speedup \\
\midrule
animal1          & $5{,}126$   & $\mathbf{0.023}$ & $0.035$ & $0.009$ & $0.014$ & $0.023$ & $1.0\times$ & $0.019$ & $1.2\times$ \\
lidense          & $12{,}509$  & $\mathbf{0.056}$ & $0.120$ & $0.019$ & $0.034$ & $0.053$ & $1.1\times$ & $0.037$ & $1.5\times$ \\
lgm\_48600\_bw2  & $48{,}600$  & $\mathbf{0.723}$ & $1.20$  & $0.163$ & $0.162$ & $0.325$ & $2.2\times$ & $0.245$ & $3.0\times$ \\
animal2          & $50{,}126$  & $16.6$ & $\mathbf{12.5}$ & $0.642$ & $1.649$ & $2.291$ & $5.5\times$ & $1.994$ & $6.3\times$ \\
lgm\_100200\_bw1 & $100{,}200$ & $\mathbf{1.13}$  & $1.79$  & $0.303$ & $0.317$ & $0.620$ & $1.8\times$ & $0.471$ & $2.4\times$ \\
lgm\_100200\_bw2 & $100{,}200$ & $\mathbf{2.38}$  & $3.37$  & $0.420$ & $0.464$ & $0.884$ & $2.7\times$ & $0.683$ & $3.5\times$ \\
\bottomrule
\end{tabular}
\end{table}

The gain tracks the work per factorization. Against the faster of the two CPU nodes, which is the Intel node on every matrix but animal2, the device is at parity on the two smallest, $1.0$ and $1.1\times$ with the factor returned, and reaches $6.3\times$ on the largest. The movement cost grows with the size of the factor while the arithmetic it hides grows faster, so the device pays only once the factor is large enough. The two speedup columns differ by the return alone, on every matrix the factor kept time is the numeric time plus the initial upload, and the kept column is the one a fit reads, since the log-determinant, the solves and the selected inversion all consume the factor where it sits. As a next step, we will develop a host/device router that supports the GPU extension of the semisparse route introduced in this work.

\section{Related work}\label{sec:related}

\paragraph{Sparse direct solvers.}
The solvers we compare against descend from the
multifrontal~\cite{duff1983multifrontal} and
supernodal~\cite{demmel1999superlu} methods: multifrontal
MUMPS~\cite{amestoy2001fully} and the supernodal
PARDISO~\cite{onemkl_pardiso}, CHOLMOD~\cite{chen2008algorithm},
PaStiX~\cite{henon2002pastix}, and symPACK~\cite{jacquelin2016sympack}. Each
recovers dense substructure bottom-up from the symbolic factor, as fronts or
supernodes, and runs it through level-3 BLAS.

\paragraph{Tile algorithms and task runtimes.}
Tile algorithms~\cite{buttari2009class,agullo2009dense,kurzak2010scheduling}
impose a uniform grid and a regular DAG and were developed for dense
factorization; sTiles~\cite{abdulfattah2025stiles} carried the fixed-grid,
statically scheduled model to the arrowhead precision matrices of latent
Gaussian models. Both keep the grid uniform, using a dense block or a sparse
CSC tile at the two ends. Task-based sparse solvers instead express the
elimination DAG over general runtimes such as
StarPU~\cite{augonnet2011starpu} or PaRSEC~\cite{bosilca2013parsec}, and pay
the runtime's dynamic scheduling on every factorization. We keep the fixed
grid and the static schedule but make the per-tile representation adaptive,
which is the active-column tile. That a build-once static schedule can rival a
dynamic runtime is itself established for
Cholesky~\cite{agullo2016static}, including recent out-of-core
results~\cite{ren2025ooc}; what is new here is building the schedule once and
replaying it across the thousands of fixed-pattern refactorizations the
workload issues. A complementary line reuses the analysis across
factorizations whose pattern \emph{changes}, as in the adaptive reordering
reuse of Parth~\cite{zarebavani2025parth}. Our regime is the opposite, one
pattern factored repeatedly, so the whole schedule, not just the ordering, is
amortized.

\paragraph{Adaptive per-block representations.}
The closest idea to per-tile adaptation is block low-rank. BLR within
MUMPS~\cite{amestoy2019blr} and tile low-rank
factorizations~\cite{akbudak2017tilelowrank} compress each off-diagonal block
to a low-rank approximation, trading a controlled accuracy loss for memory and
flops. The active-column tile adapts each tile in the same spirit but
\emph{exactly}: it drops the columns the symbolic factor proves empty, with no
approximation, and yields a factor numerically equivalent up to summation
order. (Across the suite its log-determinant matches PARDISO's to a relative
$10^{-5}$, the difference one expects from summation order between two
solvers' distinct exact factorizations; between our own CPU and GPU runs of
the same route, in Section~\ref{sec:gpu}, the agreement is $10^{-14}$.) Low-rank compresses a block's numerical rank; the
active-column tile compresses its structural occupancy. And unlike a BLR mode, it
is one representation the per-matrix selector chooses among, not a fixed
setting.

\paragraph{Auto-tuning and format selection.}
Choosing a representation per input also resembles auto-tuning and
sparse-format selection, but it differs in what is selected and how.
Auto-tuners such as OSKI~\cite{vuduc2005oski} pick register and cache
blockings for a sparse kernel by run-time profiling, and format selection
chooses a storage layout (CSR, ELLPACK, and its SIMD
variants~\cite{saad2003iterative,monakov2010ellpack,kreutzer2014unified}) to
speed a single sparse matrix-vector product. Both tune one operation's kernel
or layout. Closer still, HPS Cholesky~\cite{lin2024hps} adapts inside a sparse
factorization, training a graph-convolutional model to choose supernode
relaxation parameters per matrix. Our selector instead chooses the whole
representation of the factorization, a discrete choice among the dense,
semisparse, and supernodal tilings rather than a tuning of one
representation's parameters, and it makes that choice from three inexpensive
features of the symbolic factor before any numerical work, with no profiling
pass and no trained model. The selection is thus paid once, whereas a
profiling-based tuner would repay its measurement cost across the workload's
many factorizations. At the level of the individual tile, the semisparse
storage relates to padded sparse formats and to supernode
amalgamation~\cite{ashcraft1989relaxed}, connections we detail in
Section~\ref{sec:related-storage}.

\section{Conclusion}\label{sec:conclusion}

The symmetric positive definite systems of integrated nested Laplace
approximation, and the engineering collections beside them, span a wide
structural spectrum, both across matrices and within a single matrix, and no
one representation serves all of it. Dense matrices are alike, but sparse
matrices are each sparse in their own way. The structure in which a matrix is
best stored and factored is something to be discovered from the matrix, not a
label fixed in advance.

We have shown that one tile Cholesky factorization can cover this spectrum.
Using three features read off the symbolic factor, a per-matrix selector
routes each matrix, on one fixed grid and one static, shared-memory schedule,
to a dense tiling, to the active-column tile we introduce, or to a non-uniform
supernodal tiling. The active-column tile is the new primitive: it stores each
tile according to its own fill and reaches level-3 BLAS on the regular banded
and bordered middle that had no standard representation, and the selector is
what fuses the three modes into one solver. Across $60$ SPD matrices, from
near-diagonal graphs to full dense blocks, this routing is the only
configuration first in all three structural regimes. It beats every fixed
mode by $1.6$ to $2.6\times$ and holds an aggregate $4.7\times$ over PARDISO,
the strongest competitor, a margin set by the expensive matrices that dominate
the total, at a resident factor within a few percent of PARDISO's. The
schedule and the routing are computed once per sparsity pattern and amortized
across the thousands of factorizations a single fit issues, the
workload sTiles now serves as a solver backend in R-INLA
(\url{https://r-inla.org/sTiles}) and pyINLA
(\url{https://pyinla.org/sTiles}) \cite{rue2009approximate, fattah2026pyinla}.

A broader point is that ``sparse solver'' and ``dense solver'' comprise an overly restrictive dichotomy. A solver need not commit to one structure; it can decide, per matrix,
which representation the structure calls for. The selector here is a first and
deliberately simple instance of that decision, a short tree over three routes,
and its accuracy across a broad suite suggests the idea reaches further than
the routes we implement. Its routes and hand-set thresholds are few, and
broadening them to structures beyond this suite is what would enrich the
selector further.

This decision extends naturally from representation to hardware. The
directions we leave to future work, distributed memory and out-of-core
factorization~\cite{ren2025caracal,ren2025ooc}, together with the GPU route
Section~\ref{sec:gpu} opens, are themselves branches of the same tree: read a
matrix's structure and size, then decide whether it stays on a shared-memory
node, spills out of core, spreads across a cluster, or moves to a GPU. A good route saves not only time but
energy, since work spared from the wrong structure or the wrong device is work
never done. When one sparsity pattern is factored thousands of times, it pays
to read its structure once and let the representation, and in time the
hardware, follow.

\section*{Acknowledgments}

The research reported in this publication was supported by funding from King
Abdullah University of Science and Technology (KAUST), under the Small
Translational Research Grant (sTRG) award RFS-sTRG2025-6686. The authors
acknowledge the KAUST Supercomputing Laboratory for the computational
resources of the Ibex cluster.

\newpage

\bibliographystyle{plainnat}
\bibliography{refs}

\newpage 

\appendix
\section{Per-matrix factorization times (Intel node)}\label{app:matrices}
Table~\ref{tab:per-matrix-times} lists the $60$ matrices by their index in
Figure~\ref{fig:all-matrices} and gives the best factorization time of
every solver on each, all measured on the Intel node of
Section~\ref{sec:experiments}, as a per-matrix reference for the aggregate
results there. Bold marks the fastest solver on each matrix, which is a
different count from the \emph{wins} column of Table~\ref{tab:byclass}: that
column counts the matrices on which a solver beats sTiles, whether or not it
is also the fastest of the six. A solver can therefore beat sTiles on a matrix
that a third solver wins outright, so tallying the bold entries undercounts
the wins column.
\begin{footnotesize}
\setlength{\tabcolsep}{4pt}
\begin{longtable}{rl cccccc}
\caption{Best factorization time in seconds on the \textbf{Intel} node (minimum over the swept core counts) for every matrix and solver, sorted by sTiles time to match the numbering of Figure~\ref{fig:all-matrices}. The fastest solver on each matrix is in bold; \texttt{--} marks a matrix a solver does not factor.}\\
\label{tab:per-matrix-times}\\
\toprule
\# & matrix & sTiles & PARDISO & MUMPS & CHOLMOD & PaStiX & symPACK \\
\midrule
\endfirsthead
\multicolumn{8}{c}{\footnotesize\tablename~\thetable{} (continued)}\\
\toprule
\# & matrix & sTiles & PARDISO & MUMPS & CHOLMOD & PaStiX & symPACK \\
\midrule
\endhead
\bottomrule
\endfoot
1 & \texttt{inla\_graph\_sem\_n2000} & \textbf{0.0005} & 0.0015 & 0.0069 & 0.0015 & 0.0054 & 0.0095 \\
2 & \texttt{inla\_graph\_sem\_n5000} & \textbf{0.0008} & 0.0035 & 0.015 & 0.0041 & 0.015 & 0.020 \\
3 & \texttt{inla\_graph\_8rtKSK} & 0.0010 & 0.0011 & 0.0058 & \textbf{0.0009} & 0.078 & 3.00 \\
4 & \texttt{nasa4704} & 0.0013 & \textbf{0.0012} & 0.0058 & 0.0069 & 0.0082 & 0.0097 \\
5 & \texttt{gyro\_m} & \textbf{0.0013} & 0.0013 & 0.013 & 0.015 & 0.023 & 0.154 \\
6 & \texttt{bcsstk15} & 0.0016 & \textbf{0.0015} & 0.0096 & 0.010 & 0.0078 & 0.013 \\
7 & \texttt{gyro\_k} & \textbf{0.0018} & 0.0023 & 0.025 & 0.031 & 0.057 & 0.014 \\
8 & \texttt{inla\_graph\_sh7Pgi} & 0.0026 & \textbf{0.0019} & 0.0086 & 0.0038 & 0.0093 & 23.0 \\
9 & \texttt{inla\_graph\_pedigree} & \textbf{0.0029} & 0.0070 & 0.082 & 0.0056 & 0.041 & -- \\
10 & \texttt{inla\_graph\_sem\_n20000} & \textbf{0.0031} & 0.013 & 0.054 & 0.021 & 0.063 & 0.080 \\
11 & \texttt{inla\_graph\_diff} & \textbf{0.0032} & 0.0051 & 0.018 & 0.048 & 0.029 & 0.034 \\
12 & \texttt{msc10848} & \textbf{0.0036} & 0.0039 & 0.022 & 0.033 & 0.071 & 0.017 \\
13 & \texttt{inla\_graph\_net814381} & \textbf{0.0038} & 0.0063 & 0.032 & 0.013 & 0.031 & 1.21 \\
14 & \texttt{inla\_graph\_ayaLRw} & \textbf{0.0038} & 0.0059 & 0.027 & 0.014 & 0.029 & 0.228 \\
15 & \texttt{msc23052} & \textbf{0.0047} & 0.0050 & 0.032 & 0.051 & 0.072 & 0.021 \\
16 & \texttt{thermal1} & \textbf{0.0052} & 0.0068 & 0.036 & 0.111 & 0.043 & 0.021 \\
17 & \texttt{inla\_graph\_lgm\_10010\_bw2} & \textbf{0.010} & 0.013 & 0.036 & 0.044 & 0.048 & 0.023 \\
18 & \texttt{nasasrb} & \textbf{0.012} & 0.014 & 0.061 & 0.172 & 0.163 & 0.037 \\
19 & \texttt{oilpan} & 0.013 & \textbf{0.012} & 0.065 & 0.160 & 0.219 & 0.050 \\
20 & \texttt{inla\_graph\_sem\_n100000} & \textbf{0.014} & 0.068 & 0.276 & 0.162 & 0.353 & 0.400 \\
21 & \texttt{thermomech\_dM} & \textbf{0.015} & 0.017 & 0.090 & 0.287 & 0.125 & 0.029 \\
22 & \texttt{ct20stif} & \textbf{0.018} & 0.020 & 0.073 & 0.152 & 0.191 & 0.069 \\
23 & \texttt{s3dkt3m2} & \textbf{0.019} & 0.021 & 0.104 & 0.231 & 0.248 & 0.103 \\
24 & \texttt{smt} & \textbf{0.021} & 0.022 & 0.092 & 0.140 & 0.253 & 0.095 \\
25 & \texttt{inla\_graph\_ferris} & 0.021 & \textbf{0.013} & 0.036 & 0.143 & 0.116 & 0.087 \\
26 & \texttt{s3dkq4m2} & 0.022 & \textbf{0.021} & 0.111 & 0.247 & 0.291 & 0.101 \\
27 & \texttt{inla\_graph\_animal1} & \textbf{0.023} & 0.181 & 0.133 & 0.110 & 0.391 & 0.613 \\
28 & \texttt{apache1} & 0.024 & \textbf{0.023} & 0.081 & 0.273 & 0.079 & 0.127 \\
29 & \texttt{inla\_graph\_bern\_spd} & \textbf{0.029} & 0.044 & 0.126 & 0.511 & 0.306 & 0.775 \\
30 & \texttt{bmw7st\_1} & \textbf{0.030} & 0.033 & 0.126 & 0.394 & 0.460 & 0.175 \\
31 & \texttt{inla\_graph\_83o4NNNo} & 0.031 & \textbf{0.0082} & 0.053 & 0.120 & 0.096 & 0.023 \\
32 & \texttt{inla\_graph\_stcov} & \textbf{0.038} & 0.126 & 0.827 & 1.03 & 1.000 & 1.39 \\
33 & \texttt{inla\_graph\_spacetime} & \textbf{0.047} & 0.089 & 0.133 & 0.188 & 0.143 & 0.208 \\
34 & \texttt{nd3k} & \textbf{0.051} & 0.081 & 0.181 & 0.165 & 0.281 & 0.375 \\
35 & \texttt{pwtk} & \textbf{0.052} & 0.055 & 0.260 & 0.672 & 0.673 & 0.139 \\
36 & \texttt{inla\_graph\_lidense} & \textbf{0.056} & 0.111 & 0.377 & 0.496 & 1.55 & 1.15 \\
37 & \texttt{crankseg\_1} & 0.062 & \textbf{0.057} & 0.247 & 0.395 & 0.695 & 0.158 \\
38 & \texttt{bmw3\_2} & 0.074 & \textbf{0.070} & 0.232 & 0.685 & 0.732 & 0.320 \\
39 & \texttt{boneS01} & 0.084 & \textbf{0.083} & 0.227 & 0.533 & 0.546 & 0.552 \\
40 & \texttt{crankseg\_2} & 0.086 & \textbf{0.081} & 0.295 & 0.524 & 0.921 & 0.230 \\
41 & \texttt{tmt\_sym} & \textbf{0.087} & 0.089 & 0.254 & 1.20 & 0.453 & 0.176 \\
42 & \texttt{ecology2} & \textbf{0.107} & 0.128 & 0.326 & 1.53 & 0.485 & 0.359 \\
43 & \texttt{af\_shell3} & 0.160 & \textbf{0.152} & 0.432 & 1.37 & 1.19 & 0.409 \\
44 & \texttt{consph} & \textbf{0.188} & 0.206 & 0.512 & 0.760 & 0.719 & 1.81 \\
45 & \texttt{nd6k} & \textbf{0.208} & 0.312 & 0.475 & 0.486 & 0.679 & 2.86 \\
46 & \texttt{inline\_1} & \textbf{0.297} & 0.340 & 0.826 & 2.31 & 2.54 & 0.481 \\
47 & \texttt{inla\_graph\_yU0G1u} & 0.298 & \textbf{0.284} & 0.712 & 1.55 & 311 & -- \\
48 & \texttt{inla\_graph\_net1628760} & 0.362 & \textbf{0.229} & 0.985 & 3.70 & 3.52 & 0.819 \\
49 & \texttt{boneS10} & \textbf{0.608} & 0.629 & 1.27 & 3.69 & 3.92 & 1.13 \\
50 & \texttt{inla\_graph\_lgm\_48600\_bw2} & \textbf{0.723} & 9.43 & 4.34 & 2.36 & 8.10 & 9.45 \\
51 & \texttt{nd12k} & \textbf{0.891} & 1.47 & 1.25 & 1.48 & 1.83 & 9.27 \\
52 & \texttt{inla\_graph\_lgm\_100200\_bw1} & \textbf{1.13} & 2.48 & 1.98 & 2.22 & 4.65 & 1.85 \\
53 & \texttt{inla\_graph\_lgm\_50000\_bw15000} & \textbf{1.44} & 5.32 & 2.59 & 5.30 & 2.81 & 28.9 \\
54 & \texttt{inla\_graph\_lgm\_100200\_bw2} & \textbf{2.38} & 7.63 & 4.58 & 3.54 & 6.73 & 8.34 \\
55 & \texttt{bone010} & \textbf{6.44} & 7.88 & 7.03 & 11.8 & 9.56 & 21.8 \\
56 & \texttt{audikw\_1} & \textbf{6.67} & 12.2 & 9.18 & 14.7 & 10.5 & 47.3 \\
57 & \texttt{Fault\_639} & \textbf{7.88} & 17.1 & 10.1 & 12.4 & 10.9 & 110 \\
58 & \texttt{inla\_graph\_lgm\_50400\_bw2} & 11.9 & 37.1 & 33.5 & \textbf{7.16} & 14.9 & 76.7 \\
59 & \texttt{Emilia\_923} & \textbf{13.3} & 24.8 & 15.7 & 19.9 & 19.8 & 115 \\
60 & \texttt{inla\_graph\_animal2} & \textbf{16.6} & 209 & 35.7 & 25.0 & 72.1 & 432 \\
\end{longtable}
\end{footnotesize}

\section{Per-matrix factorization times (AMD EPYC node)}\label{app:matrices-amd}
Table~\ref{tab:per-matrix-times-amd} repeats the per-matrix reference on the
AMD EPYC node of Section~\ref{sec:setup}, over cores $\{1,\dots,128\}$. The
routing is identical to the Intel run; matrices are re-sorted by their AMD
sTiles time, and the bolding follows the same convention as
Appendix~\ref{app:matrices}.
\begin{footnotesize}
\setlength{\tabcolsep}{4pt}
\begin{longtable}{rl cccccc}
\caption{Best factorization time in seconds on the \textbf{AMD}~EPYC node (minimum over cores $\{1,\dots,128\}$) for every matrix and solver, sorted by sTiles time. The fastest solver on each matrix is in bold; \texttt{--} marks a matrix a solver does not factor.}\\
\label{tab:per-matrix-times-amd}\\
\toprule
\# & matrix & sTiles & PARDISO & MUMPS & CHOLMOD & PaStiX & symPACK \\
\midrule
\endfirsthead
\multicolumn{8}{c}{\footnotesize\tablename~\thetable{} (continued)}\\
\toprule
\# & matrix & sTiles & PARDISO & MUMPS & CHOLMOD & PaStiX & symPACK \\
\midrule
\endhead
\bottomrule
\endfoot
1 & \texttt{inla\_graph\_sem\_n2000} & \textbf{0.0004} & 0.0035 & 0.013 & 0.0013 & 0.0061 & 0.0066 \\
2 & \texttt{inla\_graph\_sem\_n5000} & \textbf{0.0009} & 0.0087 & 0.028 & 0.0033 & 0.016 & 0.015 \\
3 & \texttt{gyro\_m} & \textbf{0.0016} & 0.0051 & 0.031 & 0.014 & 0.027 & 0.194 \\
4 & \texttt{inla\_graph\_8rtKSK} & 0.0020 & 0.0023 & 0.022 & \textbf{0.0008} & 0.0049 & 4.33 \\
5 & \texttt{nasa4704} & \textbf{0.0021} & 0.0035 & 0.013 & 0.0083 & 0.0099 & 0.0056 \\
6 & \texttt{inla\_graph\_sem\_n20000} & \textbf{0.0031} & 0.029 & 0.077 & 0.020 & 0.076 & 0.061 \\
7 & \texttt{bcsstk15} & \textbf{0.0031} & 0.0053 & 0.027 & 0.012 & 0.011 & 0.0073 \\
8 & \texttt{inla\_graph\_sh7Pgi} & \textbf{0.0032} & 0.0048 & 0.025 & 0.0034 & 0.0093 & 33.7 \\
9 & \texttt{gyro\_k} & \textbf{0.0037} & 0.0083 & 0.039 & 0.035 & 0.071 & 0.0064 \\
10 & \texttt{inla\_graph\_pedigree} & \textbf{0.0041} & 0.015 & 0.073 & 0.0052 & 0.059 & -- \\
11 & \texttt{inla\_graph\_ayaLRw} & \textbf{0.0048} & 0.016 & 0.048 & 0.012 & 0.028 & 0.278 \\
12 & \texttt{msc10848} & \textbf{0.0056} & 0.012 & 0.047 & 0.041 & 0.083 & 0.014 \\
13 & \texttt{inla\_graph\_net814381} & \textbf{0.0063} & 0.017 & 0.054 & 0.012 & 0.032 & 1.61 \\
14 & \texttt{thermal1} & \textbf{0.0064} & 0.016 & 0.051 & 0.110 & 0.051 & 0.013 \\
15 & \texttt{inla\_graph\_diff} & \textbf{0.0072} & 0.0097 & 0.035 & 0.047 & 0.034 & 0.021 \\
16 & \texttt{msc23052} & \textbf{0.0078} & 0.017 & 0.051 & 0.065 & 0.079 & 0.014 \\
17 & \texttt{inla\_graph\_sem\_n100000} & \textbf{0.011} & 0.092 & 0.298 & 0.146 & 0.426 & 0.315 \\
18 & \texttt{thermomech\_dM} & 0.023 & 0.033 & 0.096 & 0.303 & 0.158 & \textbf{0.015} \\
19 & \texttt{inla\_graph\_lgm\_10010\_bw2} & 0.024 & 0.027 & 0.068 & 0.054 & 0.052 & \textbf{0.017} \\
20 & \texttt{nasasrb} & \textbf{0.024} & 0.035 & 0.113 & 0.232 & 0.176 & 0.032 \\
21 & \texttt{inla\_graph\_ferris} & 0.026 & \textbf{0.023} & 0.060 & 0.162 & 0.125 & 0.068 \\
22 & \texttt{oilpan} & \textbf{0.030} & 0.031 & 0.107 & 0.235 & 0.214 & 0.043 \\
23 & \texttt{inla\_graph\_animal1} & \textbf{0.035} & 0.260 & 0.221 & 0.167 & 0.483 & 0.939 \\
24 & \texttt{ct20stif} & \textbf{0.038} & 0.055 & 0.135 & 0.206 & 0.228 & 0.065 \\
25 & \texttt{smt} & \textbf{0.040} & 0.054 & 0.164 & 0.203 & 0.273 & 0.098 \\
26 & \texttt{apache1} & \textbf{0.041} & 0.050 & 0.162 & 0.330 & 0.094 & 0.124 \\
27 & \texttt{s3dkt3m2} & \textbf{0.043} & 0.058 & 0.171 & 0.341 & 0.245 & 0.108 \\
28 & \texttt{inla\_graph\_bern\_spd} & \textbf{0.045} & 0.053 & 0.161 & 0.540 & 0.362 & 0.719 \\
29 & \texttt{s3dkq4m2} & \textbf{0.048} & 0.063 & 0.171 & 0.350 & 0.277 & 0.107 \\
30 & \texttt{inla\_graph\_83o4NNNo} & 0.059 & 0.021 & 0.083 & 0.158 & 0.111 & \textbf{0.019} \\
31 & \texttt{inla\_graph\_stcov} & \textbf{0.064} & 0.138 & 0.977 & 0.848 & 1.00 & 0.856 \\
32 & \texttt{inla\_graph\_spacetime} & \textbf{0.069} & 0.135 & 0.243 & 0.303 & 0.187 & 0.255 \\
33 & \texttt{bmw7st\_1} & \textbf{0.073} & 0.075 & 0.203 & 0.546 & 0.469 & 0.188 \\
34 & \texttt{nd3k} & \textbf{0.087} & 0.135 & 0.325 & 0.269 & 0.347 & 0.552 \\
35 & \texttt{pwtk} & \textbf{0.113} & 0.114 & 0.382 & 0.910 & 0.649 & 0.159 \\
36 & \texttt{crankseg\_1} & \textbf{0.116} & 0.131 & 0.403 & 0.654 & 0.714 & 0.193 \\
37 & \texttt{inla\_graph\_lidense} & \textbf{0.120} & 0.251 & 0.375 & 0.408 & 1.60 & 1.80 \\
38 & \texttt{bmw3\_2} & \textbf{0.136} & 0.169 & 0.337 & 0.898 & 0.740 & 0.343 \\
39 & \texttt{tmt\_sym} & \textbf{0.137} & 0.192 & 0.311 & 1.34 & 0.414 & 0.169 \\
40 & \texttt{crankseg\_2} & \textbf{0.145} & 0.157 & 0.543 & 0.893 & 0.947 & 0.283 \\
41 & \texttt{boneS01} & \textbf{0.153} & 0.168 & 0.396 & 0.842 & 0.620 & 0.671 \\
42 & \texttt{ecology2} & \textbf{0.162} & 0.290 & 0.443 & 1.96 & 0.456 & 0.308 \\
43 & \texttt{inla\_graph\_yU0G1u} & \textbf{0.244} & 0.292 & 0.722 & 2.31 & 111 & -- \\
44 & \texttt{af\_shell3} & 0.292 & \textbf{0.282} & 0.674 & 1.98 & 1.16 & 0.429 \\
45 & \texttt{nd6k} & \textbf{0.297} & 0.364 & 0.914 & 0.848 & 0.900 & 3.94 \\
46 & \texttt{consph} & \textbf{0.330} & 0.333 & 0.993 & 1.32 & 0.877 & 2.35 \\
47 & \texttt{inline\_1} & \textbf{0.380} & 0.557 & 1.37 & 3.35 & 2.52 & 0.544 \\
48 & \texttt{inla\_graph\_net1628760} & 0.412 & 0.481 & 0.818 & 3.73 & 3.57 & \textbf{0.362} \\
49 & \texttt{boneS10} & \textbf{0.902} & 0.920 & 1.85 & 5.72 & 4.02 & 1.27 \\
50 & \texttt{inla\_graph\_lgm\_48600\_bw2} & \textbf{1.20} & 4.70 & 7.08 & 4.64 & 9.68 & 15.4 \\
51 & \texttt{nd12k} & 1.51 & \textbf{1.40} & 2.28 & 2.54 & 2.60 & 13.8 \\
52 & \texttt{inla\_graph\_lgm\_100200\_bw1} & 1.79 & \textbf{1.42} & 3.33 & 3.70 & 4.42 & 2.39 \\
53 & \texttt{inla\_graph\_lgm\_50000\_bw15000} & \textbf{2.17} & 4.60 & 4.26 & 5.47 & 3.93 & 41.6 \\
54 & \texttt{inla\_graph\_lgm\_100200\_bw2} & \textbf{3.37} & 4.15 & 7.85 & 7.11 & 6.67 & 12.2 \\
55 & \texttt{bone010} & \textbf{5.13} & 5.47 & 12.8 & 22.8 & 10.7 & 34.5 \\
56 & \texttt{audikw\_1} & \textbf{6.13} & 8.40 & 16.1 & 29.0 & 9.73 & 63.2 \\
57 & \texttt{Fault\_639} & \textbf{7.81} & 14.1 & 19.0 & 25.6 & 11.6 & 154 \\
58 & \texttt{inla\_graph\_animal2} & \textbf{12.5} & 155 & 53.8 & 30.7 & 54.6 & -- \\
59 & \texttt{Emilia\_923} & \textbf{15.3} & 20.7 & 29.7 & 40.6 & 17.1 & 163 \\
60 & \texttt{inla\_graph\_lgm\_50400\_bw2} & 22.0 & 25.8 & 50.5 & \textbf{14.8} & 15.8 & 157 \\
\end{longtable}
\end{footnotesize}

\section{Storage against an exact sparse format}\label{app:storage}
Section~\ref{sec:related-storage} compares the active-column tile with dense
tile storage. The comparison against an exact sparse format is the one that
bounds its overhead: a selected active column is stored at full height, so
zeros inside it are materialized. Table~\ref{tab:storage} measures both, over
the matrices the selector routes to this format. \emph{entries} is stored
elements over $\texttt{nnz}(L)$, the direct overhead. \emph{bytes} compares
against CSC, which pays $12$ bytes per nonzero (an fp64 value and an
\texttt{int32} row index) where the tile pays $8$ per stored entry plus one
index per active column, amortized over the tile height; with $64$-bit
indices the comparison improves further. The two disagree in sign: the tile
stores $1.22\times$ the entries but occupies $0.81\times$ the memory, and
remains a dense block for BLAS-3, which no sparse format offers. The
crossover sits near two-thirds occupancy within active columns; below it the
entry overhead dominates, as on the arrowhead \texttt{pedigree}.

\begin{table}[!ht]
\centering\small
\caption{Active-column tile storage against an exact sparse format and
against dense tiling, for the matrices the selector routes to this format.
\emph{entries} and \emph{bytes} are ratios to CSC; \emph{vs dense} is to a
full dense tiling of the same factor.}
\label{tab:storage}
\begin{tabular}{lrrrrr}
\toprule
matrix & $\texttt{nnz}(L)$ & stored & entries & bytes & vs dense \\
\midrule
\texttt{sem\_n2000} & $84{,}021$ & $96{,}000$ & $1.14\times$ & $0.76\times$ & $0.17\times$ \\
\texttt{sem\_n5000} & $210{,}021$ & $240{,}000$ & $1.14\times$ & $0.76\times$ & $0.17\times$ \\
\texttt{pedigree} & $295{,}077$ & $1{,}244{,}040$ & $4.22\times$ & $2.81\times$ & $0.25\times$ \\
\texttt{sem\_n20000} & $840{,}021$ & $960{,}000$ & $1.14\times$ & $0.76\times$ & $0.17\times$ \\
\texttt{sem\_n100000} & $4{,}200{,}021$ & $4{,}800{,}000$ & $1.14\times$ & $0.76\times$ & $0.17\times$ \\
\texttt{bern\_spd} & $4{,}326{,}417$ & $5{,}846{,}200$ & $1.35\times$ & $0.90\times$ & $0.60\times$ \\
\texttt{stcov} & $13{,}080{,}265$ & $14{,}933{,}240$ & $1.14\times$ & $0.76\times$ & $0.94\times$ \\
\midrule
aggregate & $23{,}035{,}843$ & $28{,}119{,}480$ & $1.22\times$ & $0.81\times$ & $0.42\times$ \\
\bottomrule
\end{tabular}
\end{table}

\end{document}